\documentclass{aa}
\usepackage {graphicx}
\def\Teff  {T$_{\mbox {\scriptsize eff}}$}

\def\vt  {$v_t$}

\begin {document}

%\thesaurus{06(08.01.1; 08.06.2; 08.05.3; 08.06.3; 08.16.3;  08.19.5)}
%\thesaurus{06
%  (08.01.1;    %abond
%   08.06.2;    % formation
%   08.05.3;    % evol
%   08.06.3;    %fund  param
%   08.16.3     % Pop II
%   08.19.5)}   %supernovae

\title {First stars VI - Abundances of  C, N, O, Li, and mixing in 
 extremely metal-poor giants. Galactic evolution of the light elements
\thanks{Based on observations obtained with the ESO VLT under ESO 
programme ID 165.N-0276(A). This work has made use of the SIMBAD database.}
}

\author {
M. Spite\inst{1}\and 
R. Cayrel\inst{1}\and
B. Plez\inst{2}\and
V. Hill\inst{1}\and
F. Spite\inst{1}\and
E. Depagne\inst{3}\and 
P. Fran\c cois\inst{1}\and
P. Bonifacio\inst{4}\and
B. Barbuy\inst{5}\and
T. Beers\inst{6}\and
J. Andersen\inst{7,8}\and
P. Molaro\inst{4}\and
B. Nordstr\"{o}m\inst{7,9}\and
F. Primas\inst{10}
}

\offprints {monique.spite@obspm.fr}

\institute {
  GEPI, Observatoire de Paris-Meudon, F-92125 Meudon Cedex, France,
\and 
  GRAAL, Universit\'e de Montpellier II, F-34095 Montpellier Cedex 
05, France,
\and
  European Southern Observatory (ESO), 3107 Alonso de Cordova, 
  Vitacura, Casilla 19001, Santiago 19, Chile
\and
    Osservatorio Astronomico di Trieste, INAF,
    Via G.B. Tiepolo 11, I-34131 Trieste, Italy,
\and
    IAG, Universidade de Sao Paulo, Depto. de Astronomia, 
    rua do Matao 1226, Sao Paulo 05508-900, Brazil, 
\and
Department of Physics \& Astronomy and JINA: Joint Institute for Nuclear
Astrophysics, Michigan State University, East Lansing, MI 48824, USA
\and
  Astronomical Observatory, NBIfAFG, Juliane Maries Vej 30, DK-2100 
Copenhagen,  Denmark,
\and
  Nordic Optical Telescope Scientific Association, Apartado 474, 
ES-38 700 Santa Cruz de La Palma, Spain,
\and
  Lund Observatory, Box 43, SE-221 00 Lund, Sweden,
\and
  European Southern Observatory, Karl Schwarzschild-Str. 2, 
D-85749 Garching b. M\"unchen, Germany
}

\date {Received XXX; accepted XXX}
\titlerunning {Light-element abundances and mixing in extremely 
metal-poor giants}
\authorrunning {M. Spite et al.}

%%%% ABSTRACT %%%
\abstract{
We have investigated the poorly-understood origin of nitrogen in the early 
Galaxy by determining N abundances from the NH band at 336 nm in 35 extremely 
metal-poor halo giants, with carbon and oxygen abundances from Cayrel et al. 
(2004), using high-quality ESO VLT/UVES spectra (30 of our 35 stars
are in the range ${\rm -4.1<[Fe/H]<-2.7}$ and 22 stars have 
${\rm[Fe/H]<-3.0}$).
N abundances derived both from the NH band and from the CN band at 389 nm 
for 10 stars correlate well, but show a systematic difference of 0.4~dex, 
which we attribute to uncertainties in the physical parameters of the NH band 
(line positions, gf values, dissociation energy, etc.).
Because any dredge-up of CNO processed material to the surface may
complicate the interpretation of CNO abundances in giants, we have
also measured the surface abundance of lithium in our stars as a
diagnostic of such mixing.  \\
Our sample shows a clear dichotomy between two groups of stars.  The
first group shows evidence of C to N conversion through CN cycling and
strong Li dilution, a signature of mixing; these stars are generally
more evolved and located on the upper Red Giant Branch (RGB) or
Horizontal Branch (HB).  The second group  has ${\rm
[N/Fe] <0.5}$, shows no evidence for C to N conversion, and Li is only
moderately diluted; these stars belong to the lower RGB
and we conclude that their C
and N abundances   are very close to those of the gas from which
they formed in the early Galaxy, they are called "unmixed stars".  
The [O/Fe] and [(C+N)/Fe] ratios are
the same in the two groups, confirming that the differences between
them are caused by dredge-up of CN-processed material in the first
group, with negligible contributions from the O-N cycle.\\
The "unmixed" stars reflect the abundances in the early Galaxy: the
[C/Fe] ratio is constant (about +0.2 dex) and the [C/Mg] ratio
is close to solar at low metallicity, favouring a high C production 
by massive zero-metal supernovae.  The [N/Fe] and [N/Mg] ratios 
scatter widely.  Their mean values in each metallicity bin decrease
with increasing metallicity, but this trend could be well a
statistical effect.  The larger values of these ratios define a flat
upper plateau ([N/Mg]= 0.0, [N/Fe]= +0.1), which could reflect higher values 
within a wide range of yields of zero-metal SNe~II. Alternatively, by 
analogy with the DLA's, the lower abundances ([N/Mg]= --1.1, [N/Fe]= --0.7) 
could reflect generally low yields from the first SNe~II, the other 
stars being N enhanced by winds of massive Asymptotic Giant Branch 
(AGB) stars. Since all the stars show clear [$\alpha$/Fe] 
enhancements, they were formed before any significant enrichment of the 
Galactic gas by SNe~Ia, and their composition should reflect the yields of 
the first SNe II. However, if massive AGB stars
or AGB supernovae evolved more rapidly than SNe Ia and contaminated
the ISM, our stars would also reflect the yields of these
AGB stars.  At present it cannot be decided whether primary N is
produced primarily in SNe II or in massive AGB stars, or in both.\\
The stellar N abundances and [N/O] ratios are compatible with those
found in Damped Lyman-$\alpha$ (DLA) systems.  They extend the
well-known DLA ``plateau'' at [N/O] $\approx -0.8$ to lower
metallicities, albeit with more scatter; no star is found below the
putative ``low [N/$\alpha$] plateau'' at [N/O]$\approx -1.55$ in DLAs.
\keywords {Galaxy: abundances -- Galaxy: halo -- Galaxy: evolution -- 
Stars: abundances -- Stars: Mixing -- Stars: Supernovae}
}
\maketitle
%
%________________________________________________________________
%%%% 1-INTRODUCTION %%%
\section {Introduction}

% FIGURE 1
\begin {figure*}
\begin {center}
\resizebox  {12.cm}{6.0cm} 
{\includegraphics {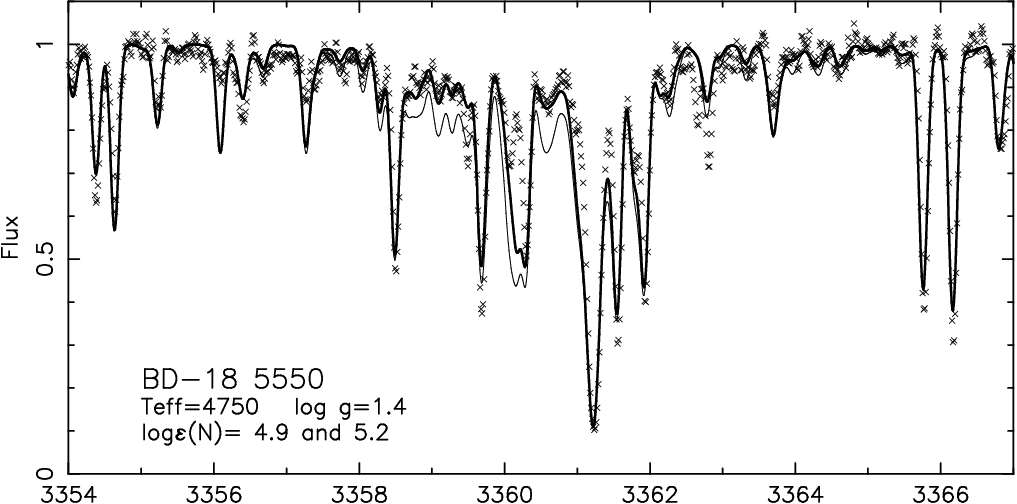} }
\resizebox  {12.cm}{6.0cm} 
{\includegraphics {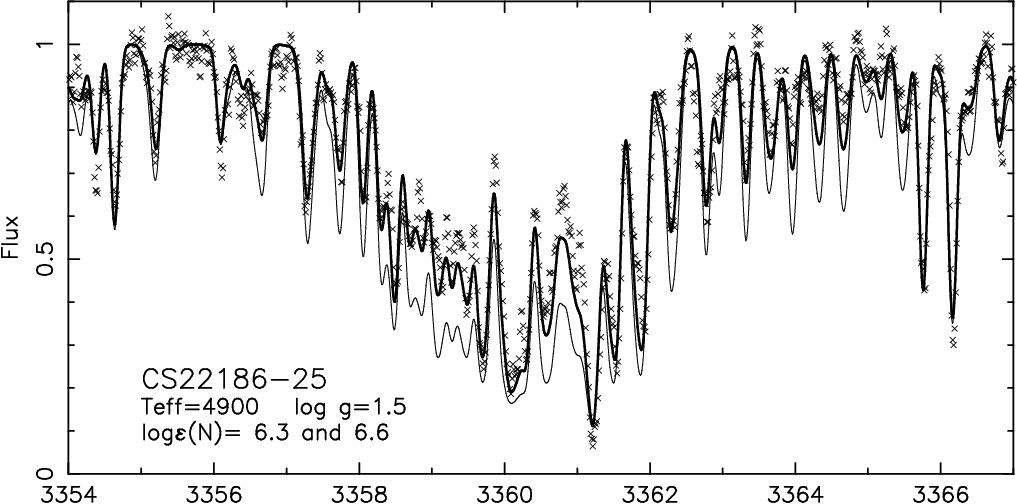} }
\caption {Reduced UVES spectra in the region of the NH band at 336 nm. Crosses 
represent the observed spectra, while the full lines show synthetic spectra 
computed for the best-fit N abundance as well as twice that value (thick and 
thin lines, respectively). The two stars have about the same metallicity, 
[Fe/H]= --3.06 (BD-18:5550) and [Fe/H]= --3.00 (CS~22186--025).}
\label {spectra}
\end {center}
\end {figure*}

Carbon, nitrogen, and oxygen - the CNO elements -- are key players in the 
chemical evolution of galaxies. They are the most abundant elements after 
hydrogen and helium and the most efficient coolants in the interstellar medium. 
Thus, understanding how the first CNO nuclei were synthesised is crucial for 
models of the first star formation and nucleosynthesis in the Universe after the 
Big Bang.

There is general consensus that oxygen is almost entirely contributed by massive 
type II SNe, primarily during the central hydrogen burning with some 
contribution from neon burning. Carbon can be produced in stars of all masses, 
essentially by He burning. In contrast to O and C, however, the initial 
formation of N is still not well understood (Pilyugin et al. \cite{PTV03}). 

N is formed at the expense of C and O during hydrogen burning by the CNO cycle. 
If N is formed directly from He, it is called primary; if pre-existing C nuclei 
are required, it is called secondary. In massive stars, mixing (with or without 
stellar rotation) between the helium-burning layer (which produces C) and the 
hydrogen-burning layer can induce ``primary'' N formation; the yield may be very
variable. In intermediate-mass stars with (approx. $4 M_{\odot} < M <8 
M_{\odot} $), abundant primary N may be produced during the AGB phase. 

The production mechanisms of N are imprinted in the abundances of stars formed 
from the ejecta of massive progenitor stars. If the N production was secondary, 
the [N/O] ratio should increase with increasing metallicity. For primary N 
production, [N/O] should remain constant.

From a study of H~II regions in irregular galaxies (Kobulnicky \& Skillman
\cite{KS96}; Izotov \& Thuan \cite{IT99}), the [N/O] ratio does indeed appear to 
be constant at low metallicity, at least between [O/H]= --1.6 and --1.2, or
about $-2.1 < {\rm [Fe/H]} < -1.7$, suggesting that N is primary in this range. 
From the small dispersion of their [N/O] measurements, Izotov \& Thuan argue in 
favor of N production by massive stars. However, this conclusion is challenged 
by observations of DLAs, which exhibit large scatter and, in particular, lower 
values of [N/O] than those of H~II regions in irregular galaxies. Henry et al. 
(\cite{HEK00}) could reproduce the plateau in [N/O] by assuming N production by 
AGB stars and different Star Formation Rates (SFRs), while Prochaska et al. 
(\cite{PHO02}) proposed a scenario involving a different IMF at earlier epochs 
of star formation. The complex evidence suggests that primary production of N, 
if confirmed, might require production by both massive stars or AGBs.

Recently, however, Chiappini, Matteucci \& Meynet (\cite{CMM03}) questioned the 
very presence of the plateau and its significance as an evolutionary curve. They 
further stressed that while the [N/O] vs. [O/H] diagram for DLAs and H~II 
regions is often interpreted as an evolutionary diagram with [O/H] as the time 
axis, it does in fact represent final abundances achieved by objects that 
evolved in completely different ways from each other. Hence, the best way to 
determine the lower limit of the [N/O] ratio is to measure it in very old 
Galactic stars. 

Among early discussions of the Galactic evolution of N in halo stars, Tomkin \& 
Lambert (\cite {TL84}) analysed 14 disk and halo dwarfs in the range ${\rm -2.3 
\leq [Fe/H] \leq -0.3}$, using the ultraviolet NH band. They found [N/Fe] =
--0.25, [C/Fe] = --0.21, and an average [N/C] = -0.02$\pm$0.3. Laird (1985)
analysed 116 dwarfs with -2.45 $\leq$ [Fe/H] $\leq$ +0.5 from observations of 
the G band of CH and the violet NH band at low resolution and found ${\rm [C/Fe] 
= -0.22\pm 0.14}$ and ${\rm [N/Fe] = -0.67\pm 0.21}$. In both studies, the N 
abundances were corrected by +0.65 dex and C by +0.2 dex in order to obtain 
[N/Fe] = [C/Fe] = 0 at solar metallicity. Carbon et al. (1987) also observed the 
CH band and violet NH band at low resolution in 83 dwarfs, including 27 in the 
range --2.7 $\leq$ [Fe/H] $\leq$ --2.0. Assuming [O/Fe] = +0.6 (important for 
the derived C abundance), they found [C/Fe] = --0.03$\pm$0.18, and [N/Fe] = 
--0.45$\pm$0.28. These three sets of data suggested that Nitrogen 
showed a behaviour as a primary element at the low metallicities 
studied  (--2.7$\leq$ [Fe/H] $\leq$ --2.0).

Very recently, Israelian et al. (\cite {IER04}) measured nitrogen abundances in 
31 unevolved stars with ${\rm -3.05 \leq [Fe/H] \leq -0.35}$, using the near-UV 
NH band. As only four stars in their sample have ${\rm [Fe/H] < -2.5}$, their 
metallicity range is complementary to the present work (see below). Israelian et 
al. find no trend of [N/Fe] vs. [Fe/H], suggesting also primary N production. For 
stars of higher metallicity (${\rm [O/H] > -1.8}$) they find a significant slope 
of [N/O] versus [O/H], indicative of secondary behaviour. Because they derive 
[N/O] from similar analyses of the near-UV bands of OH and NH, systematic errors 
in the molecular analysis should at least partly cancel.

On this background, the aim of the present paper is to determine C and N 
abundances in stars from the very earliest phases of the evolution of the 
Galaxy. Cayrel et al. (\cite {CDS04}, hereafter Paper V) already analysed the 
abundances of 17 elements from C to Zn in 35 extremely metal-poor (XMP) stars ($ 
-4.0 <{\rm[Fe/H]}<-2.7 $). For most elements, the diagrams of [X/Fe] (or [X/Mg]) 
vs. [Fe/H] (or [Mg/H]) showed very small dispersion, and the trends (or absence 
of trends) of the different element ratios with metallicity could be determined 
rather easily. In contrast, the relations [C/Fe] vs. [Fe/H] (or [C/Mg] vs. [Mg/H]) 
showed so large scatter that no conclusions on the production of C could be 
drawn. N abundances were measured from CN, but this band was only detected in 
six stars.

In this paper we use the high UV efficiency of the VLT spectrograph UVES to push 
the study of the Galactic evolution of N a step further. Using the NH band at 
336 nm allows us to measure N abundances in the same large sample of extremely 
metal-poor halo giants as that of Cayrel et al. (\cite{CDS04}): 35 stars with 
--4.1 $\leq$ [Fe/H] $\leq$ --2.0, 30 of which have $-4<{\rm [Fe/H]<}-2.7$ (and 
22 have ${\rm [Fe/H]}<-3.0$). At these metallicities it is essentially 
impossible to detect the NH band in main-sequence or turnoff stars (unless they 
have a strong N excess), so the less-evolved, unmixed giants in our sample offer 
essentially the only way to measure N at such low metallicities.

Mixing of the outer layers in these stars may invalidate conclusions
drawn from an abundance analysis of their surfaces.  Its importance
can be estimated from several indicators, in particular from the
abundance of lithium, which is rapidly destroyed if convection drives it
to regions with temperatures above $\sim2.5~10^6$ K. For example $^7$Li is
destroyed in about 97\% of the mass of a $0.9 M_{\odot}$ very
metal-poor star during its main-sequence evolution.  When the outer
convective zone deepens, the remaining 3\% is diluted by the full mass
of the convective zone, and the surface Li abundance becomes a
diagnostic of the depth of this convective zone.  Accordingly, we have
also measured Li abundances in our giants to distinguish between mixed and
unmixed stars.
 
In Sect. 2 we describe the observations, and Sect. 3 presents the analysis of 
the data. Sect. 4 discusses mixing in these stars, while Sect. 5 considers the 
abundances of the CNO elements in the early Galaxy and presents a comparison of 
our results with data from DLAs.  Sect. 6 is a brief summary of our results and 
conclusions.

%\begin{tabular}{l@{ }l@{ }c@{ }c@{ }c@{ } c@{ }c@{ }c@{ }c@{ }c@{ }c@{ }c@{ }c@{ }c}
%\begin {tabular}{llcccccccccccc}
% TABLE 1
\begin {table*}[t]
\caption {Adopted model parameters (\Teff, log $g$, \vt, [Fe/H])
and light element abundances for the programme stars. For Li, C and N the 
1$\sigma$ measurement error is given. The N abundances from the 
NH band in column 10 are raw values, to be corrected by -0.4 dex (see section 
\ref{azot1} and Fig. \ref {CN-NH}), the adopted values of [N/H] are 
given in column 11.    Remark ``m'' in the last column 
identifies stars considered to be mixed (see section \ref{BehavCN}).
The errors indicated in this table are only the measurement errors.}
\label {tabund}
\begin {center}
\begin{tabular}{llc@{ }c@{ }c@{ } c@{ }c@{ }c@{ }c@{ }c@{ }c@{ }c@{ }c@{ }c}
\hline \hline
&              &            &      &   &       &              &                &[N/H]   &    [N/H]      & adopt.\\
&Star          &   $T_{eff}$& log g&\vt&[Fe/H] & logN(Li)     & [C/H]          & (CN)   &    (NH)       & [N/H] &[O/H] &[${\rm C+N \over Fe}$] & R\\
\hline
~1 & HD~2796        & 4950 & 1.5 & 2.1 & -2.47 &$ <-0.30      $&$ -2.97\pm 0.06$& -1.52 &$-1.22\pm 0.08$& -1.62& -1.97&  0.14&  m\\
~2 & HD~122563      & 4600 & 1.1 & 2.0 & -2.82 &$ <-0.60      $&$ -3.29\pm 0.05$& -2.12 &$-1.72\pm 0.15$& -2.12& -2.20&  0.04&  m\\
~3 & HD~186478      & 4700 & 1.3 & 2.0 & -2.59 &$ <-0.50      $&$ -2.89\pm 0.07$& -2.02 &$-1.57\pm 0.12$& -1.97& -1.84&  0.01&  m\\
~4 & BD~+17:3248    & 5250 & 1.4 & 1.5 & -2.07 &$ < 0.00      $&$ -2.44\pm 0.05$& -1.37 &$-1.02\pm 0.10$& -1.42& -1.38&  0.00&  m\\
~5 & BD~--18:5550   & 4750 & 1.4 & 1.8 & -3.06 &$ 0.75\pm 0.04$&$ -3.08\pm 0.04$&   -   &$-3.02\pm 0.10$& -3.42& -2.64& -0.08&   \\
~6 & CD~--38:245    & 4800 & 1.5 & 2.2 & -4.19 &$<-0.40       $&$<-4.52        $&   -   &$-2.72\pm 0.20$& -3.12&   -  &$<$0.35& m\\
~7 & BS~16467--062  & 5200 & 2.5 & 1.6 & -3.77 &$ 0.80\pm 0.08$&$ -3.52\pm 0.12$&   -   &$< -2.92      $&$<-3.32$&   -  &   -  &   \\
~8 & BS~16477--003  & 4900 & 1.7 & 1.8 & -3.36 &$ 0.90\pm 0.04$&$ -3.07\pm 0.08$&   -   &$< -3.22      $&$<-3.62$& -  &   -  &   \\
~9 & BS~17569--049  & 4700 & 1.2 & 1.9 & -2.88 &$<-0.30       $&$ -3.00\pm 0.05$& -2.12 &$-1.62\pm 0.12$& -2.02& -  &  0.22&  m\\
10 & CS~22169--035  & 4700 & 1.2 & 2.2 & -3.04 &$<-0.35       $&$ -3.26\pm 0.05$& -2.02 &$-1.62\pm 0.13$& -2.02&   -  &  0.32&  m\\
11 & CS~22172--002  & 4800 & 1.3 & 2.2 & -3.86 &$ 0.32\pm 0.11$&$ -3.86\pm 0.10$&   -   &$-3.22\pm 0.20$& -3.62& -2.82&  0.01&   \\
12 & CS~22186--025  & 4900 & 1.5 & 2.0 & -3.00 &$<-0.15       $&$ -3.54\pm 0.10$&   -   &$-1.62\pm 0.08$& -2.02& -2.41&  0.23&  m\\
13 & CS~22189--009  & 4900 & 1.7 & 1.9 & -3.49 &$ 0.54\pm 0.05$&$ -3.18\pm 0.08$&   -   &$-2.82\pm 0.12$& -3.22&   -  &  0.25&   \\
14 & CS~22873--055  & 4550 & 0.7 & 2.2 & -2.99 &$<-0.35       $&$ -3.72\pm 0.06$&   -   &$-1.52\pm 0.20$& -1.92& -2.47&  0.19&  m\\
15 & CS~22873--166  & 4550 & 0.9 & 2.1 & -2.97 &$<-0.55       $&$ -3.10\pm 0.08$& -1.92 &$-1.52\pm 0.20$& -1.92&   -  &  0.37&  m\\
16 & CS~22878--101  & 4800 & 1.3 & 2.0 & -3.25 &$<-0.30       $&$ -3.54\pm 0.10$&   -   &$-1.52\pm 0.10$& -1.92&   -  &  0.58&  m\\
17 & CS~22885--096  & 5050 & 2.6 & 1.8 & -3.78 &$ 0.83\pm 0.05$&$ -3.52\pm 0.06$&   -   &$-3.12\pm 0.13$& -3.52&   -  &  0.20&   \\
18 & CS~22891--209  & 4700 & 1.0 & 2.1 & -3.29 &$<-0.50       $&$ -3.94\pm 0.05$&   -   &$-1.77\pm 0.10$& -2.17& -2.52&  0.36&  m\\
19 & CS~22892--052  & 4850 & 1.6 & 1.9 & -3.03 &$ 0.20\pm 0.09$&$ -2.14\pm 0.06$& -2.32 &$-2.12\pm 0.13$& -2.52& -2.56&  0.83&   \\
20 & CS~22896--154  & 5250 & 2.7 & 1.2 & -2.69 &$ 1.15\pm 0.04$&$ -2.46\pm 0.05$&   -   &$-2.52\pm 0.12$& -2.92& -1.75&  0.19&   \\
21 & CS~22897--008  & 4900 & 1.7 & 2.0 & -3.41 &$ 0.90\pm 0.04$&$ -2.85\pm 0.05$&   -   &$-2.77\pm 0.15$& -3.17&   -  &  0.45&   \\
22 & CS~22948--066  & 5100 & 1.8 & 2.0 & -3.14 &$< 0.00       $&$ -3.14\pm 0.10$&   -   &$-1.52\pm 0.10$& -1.92& -2.25&  0.61&  m\\
23 & CS~22949--037  & 4900 & 1.5 & 1.8 & -3.97 &$<-0.30       $&$ -2.82\pm 0.10$& -1.40 &$-1.32\pm 0.30$& -1.82& -1.99&  1.67&  m\\
24 & CS~22952--015  & 4800 & 1.3 & 2.1 & -3.43 &$<-0.40       $&$ -4.02\pm 0.08$&   -   &$-1.72\pm 0.10$& -2.12&   -  &  0.54&  m\\
25 & CS~22953--003  & 5100 & 2.3 & 1.7 & -2.84 &$ 0.80\pm 0.04$&$ -2.54\pm 0.03$&   -   &$-2.32\pm 0.10$& -2.72& -2.09&  0.57&   \\
26 & CS~22956--050  & 4900 & 1.7 & 1.8 & -3.33 &$ 0.97\pm 0.03$&$ -3.06\pm 0.05$&   -   &$-2.62\pm 0.10$& -3.02& -2.21&  0.26&   \\
27 & CS~22966--057  & 5300 & 2.2 & 1.4 & -2.62 &$ 1.35\pm 0.03$&$ -2.56\pm 0.05$&   -   &$-2.12\pm 0.12$& -2.52& -1.63&  0.04&   \\
28 & CS~22968--014  & 4850 & 1.7 & 1.9 & -3.56 &$ 0.52\pm 0.08$&$ -3.31\pm 0.06$&   -   &$-2.92\pm 0.10$& -3.32& -2.66&  0.25&   \\
29 & CS~29491--053  & 4700 & 1.3 & 2.0 & -3.04 &$<-0.10       $&$ -3.31\pm 0.05$& -2.32 &$-1.82\pm 0.15$& -2.22& -2.28&  0.16&  m\\
30 & CS~29495--041  & 4800 & 1.5 & 1.8 & -2.82 &$ 0.38\pm 0.09$&$ -2.89\pm 0.06$&   -   &$-2.02\pm 0.10$& -2.42& -2.14&  0.05&   \\
31 & CS~29502--042  & 5100 & 2.5 & 1.5 & -3.19 &$ 0.90\pm 0.05$&$ -3.03\pm 0.04$&   -   &$-3.22\pm 0.20$& -3.62&   -  &  0.15&   \\
32 & CS~29516--024  & 4650 & 1.2 & 1.7 & -3.06 &$ 0.33\pm 0.04$&$ -3.12\pm 0.05$&   -   &$-3.42\pm 0.20$& -3.82& -2.44& -0.09&   \\
33 & CS~29518--051  & 5200 & 2.6 & 1.4 & -2.69 &$<-0.10       $&$ -2.82\pm 0.05$&   -   &$-1.47\pm 0.15$& -1.87& -1.89&  0.24&  m\\
34 & CS~30325--094  & 4950 & 2.0 & 1.5 & -3.30 &$<-0.25       $&$ -3.30\pm 0.05$&   -   &$-2.72\pm 0.18$& -3.12& -2.58&  0.00&   \\
35 & CS~31082--001  & 4825 & 1.5 & 1.8 & -2.91 &$ 0.85\pm 0.05$&$ -2.69\pm 0.05$&$<$-2.70&$-3.02\pm0.10$& -3.42& -2.31&  0.13&  \\
\hline 
\end {tabular}
\end {center}
\end {table*}

%%%% 2-OBSERVATIONS ET REDUCTIONS %%%
\section {Observations and reductions}

The observations were performed during several observing runs from April 2000 to 
November 2001 with the VLT-UT2 and UVES (Dekker et al. \cite{DD00}), at a 
resolving power of $R=47,000$ at 400 nm. The spectra were reduced using the UVES 
context (Ballester et al. \cite{BMB00}) within MIDAS. Details of the observing 
and reduction procedures are presented in Paper V (Cayrel et al. \cite{CDS04}). 
Taking advantage of the high UV efficiency of UVES allows us to measure the NH 
band at 336 nm in most of our stars. 

The signal-to-noise ratio of the spectra is difficult to estimate in this very 
crowded region, but is typically $S/N \approx 30$ (per pixel). An example of the 
spectrum in the region of the NH band is shown in Fig. \ref{spectra}.

\section {Abundance Analysis}

As described in Paper V, we carried out a classical LTE analysis using OSMARCS
model atmospheres (Gustafsson et al. \cite{GBE75}; Plez et al. \cite{PBN92};
Edvardsson et al. \cite{EAG93}; Asplund et al. \cite{AGK97}; Gustafsson et al.
\cite{GEE03}). Since our analysis uses 1D models we adopted, as in Paper V, 
the solar abundances as obtained also from 1D models, i.e., log $\epsilon$
(O) = 8.74, log $\epsilon$ (C) = 8.52, log $\epsilon$ (N) = 7.92, and the values
for other elements from Grevesse \& Sauval (\cite{GS98}). 

Abundances were derived using the current version of the Turbospectrum code 
(Alvarez \& Plez \cite{AP98}), which accounts properly for continuum scattering 
(see Paper V), a feature that is particularly important in the violet part of 
the spectrum. For the CH, CN, and NH bands the abundances of C and N were 
determined from spectrum synthesis. Abundances of O from the forbidden line at 
630.031~nm and of Li from the resonance line have been derived directly from the 
equivalent widths of these lines.

\subsection {Atmospheric Parameters}

The procedures used to derive \Teff, log $g$, and the microturbulent velocity 
\vt\ have been explained in detail in Paper V, Section 3. Briefly, \Teff\ was 
derived from broadband photometry ({\it $B-V$, $V-R$, $V-I$, $V-K$, $J-K$}), 
calibrated by the IRFM method (Alonso et al. \cite {AAM99}). The log $g$ 
value was
obtained by requiring that identical Fe and Ti abundances be derived from Fe~I 
and Fe~II, or Ti~I, Ti~II lines, respectively, and \vt\ was determined to 
eliminate any abundance trend of the Fe~I lines with equivalent width.

\subsection {Abundances of the light elements}

\subsubsection {Lithium}

The Li line at 670.7~nm is visible in about half of our stars. Table 
\ref{tabund} lists the derived Li abundances for these stars, and upper limits 
for the remaining stars. 

\begin {figure}
\begin {center}
\resizebox  {8.cm}{4.0cm} 
{\includegraphics {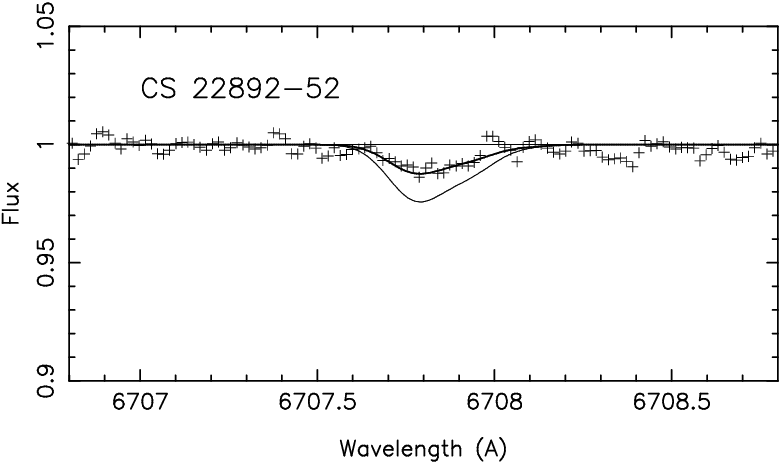} }
\caption {Observed spectrum of CS~22892-52 near the Li line at 670.7 nm 
(crosses), and synthetic spectra computed with log N(Li)= 0.15 and 0.45 (thick 
and thin lines, respectively).}
\label {Li}
\end {center}
\end {figure}

Fig. \ref {Li} shows the Li feature in the neutron-capture-rich XMP star 
CS~22892--052 (Sneden et al. \cite {SCL03}). We measure an equivalent width of 
3.3 m\AA, in excellent agreement with the value W$_{Li}$= 3.5 m\AA\ by Sneden et 
al. (\cite{SCL03}).

% FIGURE 2

\subsubsection {Carbon and Oxygen}  \label{AbCandO}

Carbon and oxygen abundances for our stars were carefully determined in Paper V 
(specifically  in sections 4.1 and 4.2) and are reproduced in Table \ref{tabund}
(see also Barbuy et al. \cite {BMS03}).

The O abundance was derived from the [O I] line at 630.031~nm (Table 
\ref{tabund}), generally considered to be the most reliable O abundance 
indicator since it is insensitive to non-LTE effects. However, these values were 
computed with classical 1D models, and the line has been shown to be sensitive 
to hydrodynamical (3D) effects: The [O/Fe] ratio based on the [O~I] line is 
expected to decrease when computed with 3D models. So far, explicit 3D 
corrections have only been computed for dwarfs (Nissen et al. \cite{NPA02}). 
Following these authors, at least the sign of the correction should not change 
in giants; and we assume, as a first approximation, that the correction is the 
same as for dwarfs. The 1D values of [O/Fe]= [O/H]--[Fe/H] (Table \ref{tabund}) 
should then be corrected by about --0.23 dex (see Paper V).  

C abundances were determined by a synthetic spectrum fit to the G band of CH 
(see Paper V).

\begin {figure}
\begin {center}
\resizebox  {8.cm}{5.45cm} 
{\includegraphics {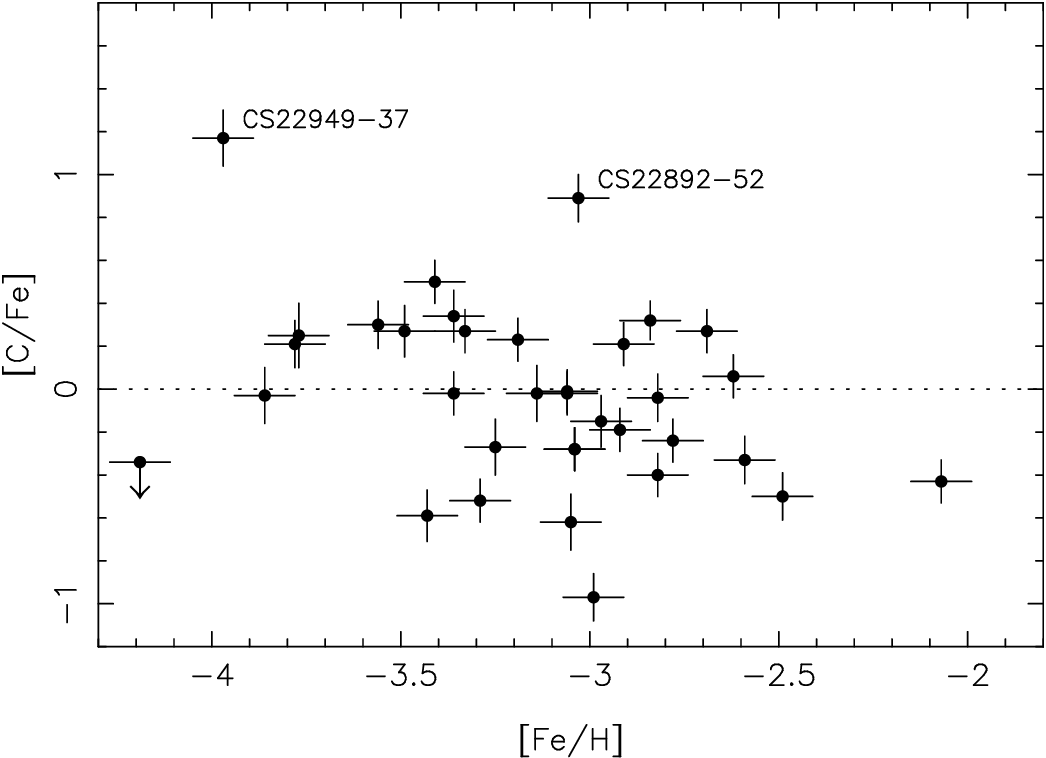} }
\resizebox  {8.cm}{0.5cm} 
{\includegraphics {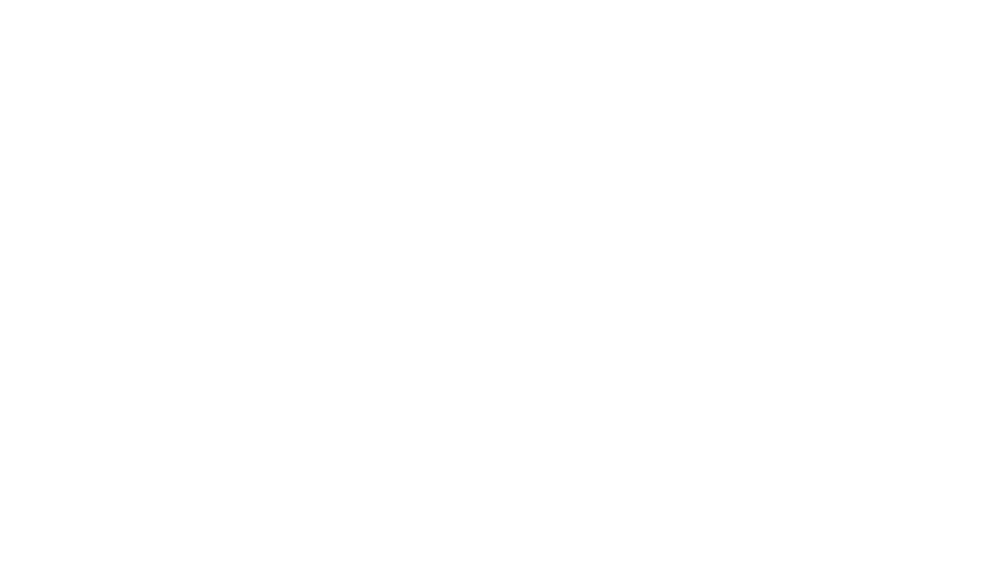} }
\resizebox  {8.cm}{5.45cm} 
{\includegraphics {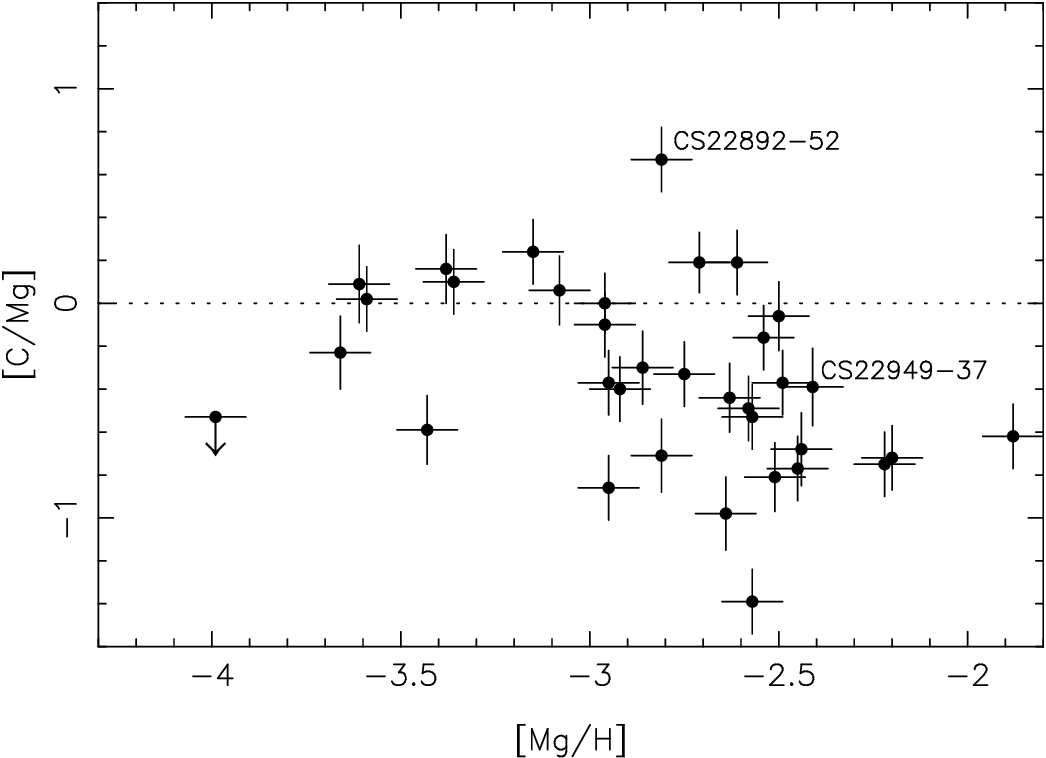} }
\caption {[C/Fe] and [C/Mg] vs. [Fe/H] and [Mg/H]. Both diagrams show far more 
scatter than seen for other elements in Paper V. The peculiar ``carbon-rich'' 
stars CS~22949--037 and CS~22892-052 are labelled; they cannot be directly 
compared to the other stars of the sample.}
\label {Carbrut}
\end {center}
\end {figure}

Fig \ref {Carbrut} shows the [C/Fe] and [C/Mg] ratios as functions of [Fe/H]
and [Mg/H]. They show substantially larger scatter around the mean value than 
all other elements studied in Paper V. Even if the two peculiar stars with very 
strong C enhancements are excluded, i.e. CS~22892--052 (Sneden et al. 
\cite{SMP96}, \cite{SCI00}, \cite{SCL03}) and CS~22949--037 (McWilliam et al. 
\cite{MPS95}, Norris et al. \cite{NRB01}, Depagne et al. \cite{DHS02}), the 
scatter remains very large (${\rm <[C/Fe]>}$ = --0.09 and $\sigma = 0.373$). In 
Paper V we studied a subset of stars with ${\rm T_{eff}} > 4800$ K (the hotter, 
less evolved part of the RGB), but the scatter remained almost constant, with 
${\rm <[C/Fe]>}$ = +0.01 and $\sigma = 0.367$ (see Figs. 5 and 12 in Paper V).

We conclude that either the gas from which the XMP stars formed had a wide range 
of C abundances, or the initial composition in the atmosphere of the stars has 
been altered by subsequent mixing episodes. In the latter case we would also 
expect a correspondingly large scatter in the relation of [N/Fe] vs. [Fe/H].

% FIGURE 3

\subsubsection {Nitrogen}  \label{azot1}

\begin {figure}
\begin {center}
\resizebox  {5.cm}{5.0cm} 
{\includegraphics {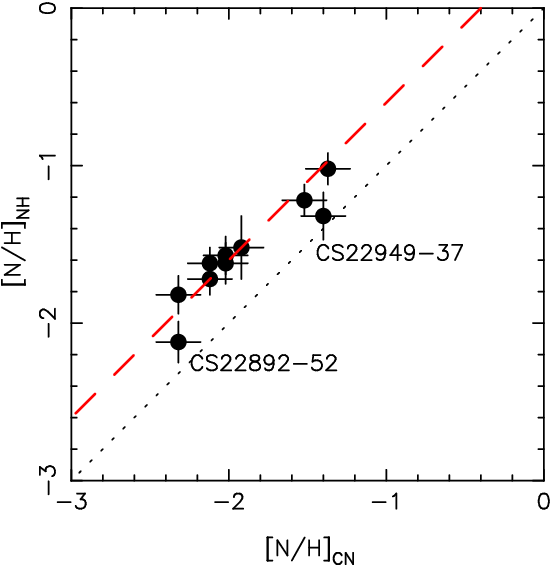} }
\caption {[N/H] values derived from the CN and NH bands. The correlation is 
good, but shows a systematic shift of about 0.4 dex. 
The two carbon-rich stars are identified.}
\label {CN-NH}
\end {center}
\end {figure}

In Paper V, N abundances were computed from the BX band of CN at 388.8 nm. This 
band is extremely weak in most of our sample and the nitrogen 
abundance had been computed in only six 
stars. After a careful examination of the spectra, we could in  fact 
detect the CN band and measure it, in ten stars (Table \ref{tabund}).\\ 
Moreover in the present paper, we use the lines of the violet 
$A^{3} \Pi_{i} - X^{3} \Sigma^{-}$ NH band at 336 nm (see Fig. \ref{spectra}), 
which is not only more readily measurable but also insensitive to the C and O 
abundances (see e.g. Sneden \cite {Sne73} or Norris et al. \cite{NRB02}). We 
have adopted the Kurucz (\cite{Kur01}) data for the NH molecule (Table 
\ref{tabund}, Fig. \ref{CN-NH}), in particular a dissociation energy of 3.47 eV 
(Huber \& Herzberg \cite {HH79}).

Fig. \ref{CN-NH} compares the N abundances derived from the NH and CN molecular 
bands. The correlation is good, but there is a systematic difference of about 
0.4 dex, well above the internal errors. The reason for this discrepancy is 
unclear, but we recall that the physical parameters (line positions, gf values, 
etc.) of the NH band are not yet well established. We use a dissociation 
energy of 3.47 eV (Huber \& Herzberg \cite{HH79}), while Sneden (\cite {Sne73}) and 
Norris et al. (\cite{NRB02}) preferred a value of 3.21 eV. However, adopting a 
lower dissociation energy would further {\it increase} our N abundances from NH 
and the discrepancy with the results from CN. In contrast, the dissociation 
energy and other parameters of the CN molecule seem to be better established.
We have adopted ${\rm D_{CN}=7.76eV}$ (Huber \& Herzberg \cite{HH79}), which is
supported by more recent experimental 
(e.g. 7.74 $\pm 0.02$, Huang et al. \cite{Huang92}) and theoretical values 
(e.g. 7.72 $\pm 0.02$, Pradhan et al. \cite{PPB94}).

We conclude that our nitrogen abundances from the NH band (Table \ref{tabund}) 
should be corrected by --0.4~dex and use these corrected values in the following 
discussions. 

\begin {figure}
\begin {center}
\resizebox  {8.cm}{5.45cm} 
{\includegraphics {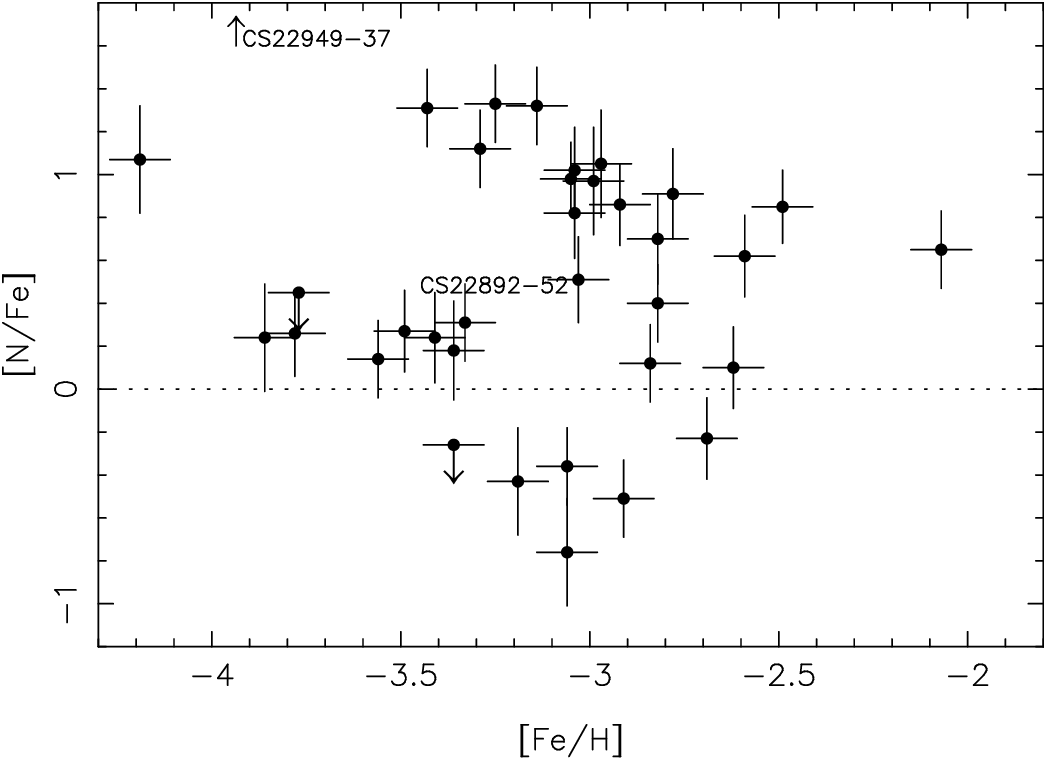} }
\resizebox  {8.cm}{0.5cm} 
{\includegraphics {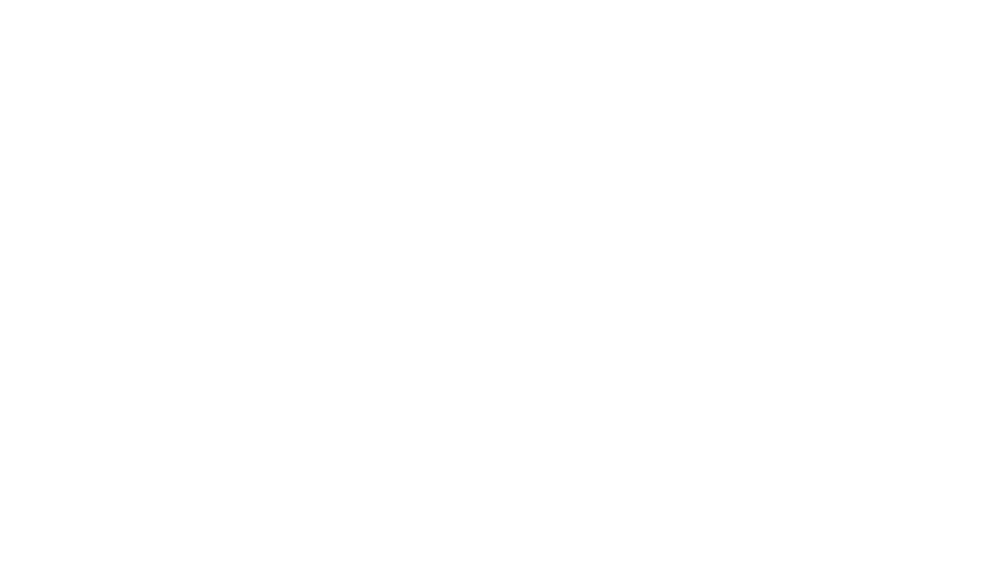} }
\resizebox  {8.cm}{5.45cm} 
{\includegraphics {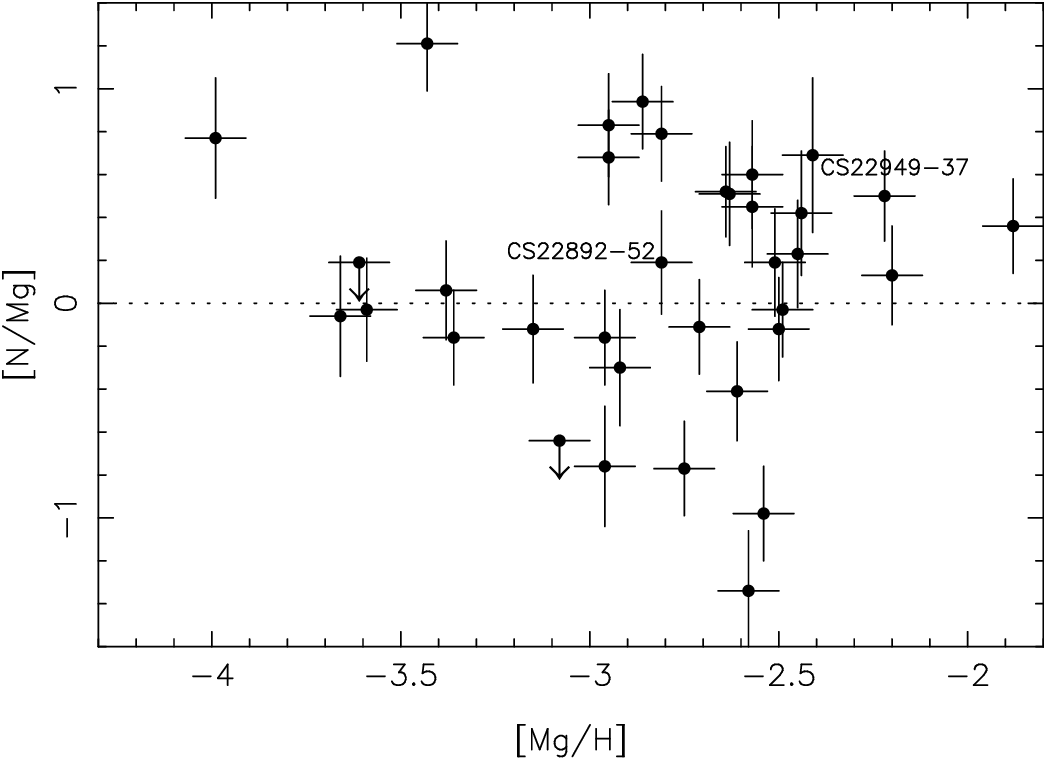} }
\caption {[N/Fe] vs. [Fe/H] and [N/Mg] vs. [Mg/H] for our sample.  The scatter 
of the points is even larger than in Fig. \ref{Carbrut}
(the scale of both figures is the same).} 
\label {Azbrut}
\end {center}
\end {figure}

Fig. \ref{Azbrut} shows the relations [N/Fe] vs. [Fe/H] and [N/Mg] vs. [Mg/H]. 
Here again the scatter is extremely large, even greater than in the case of C.

% FIGURE 4

% FIGURE 5

\subsection {Errors due to the uncertainty of the atmospheric parameters}

Except for measurement errors, the main source of error in the C,
N, and Li abundances is the uncertainty in \Teff, which is about 80 K
(Paper V).  For given \Teff, the ionization equilibrium constrains log
$g$ with an internal error of 0.1 dex, while \vt\ is constrained to
within $\pm0.2$ km~s$^{-1}$.  In Table \ref {errorsN} we list the
uncertainties in logN(Li), [C/Fe], [N/Fe] and also [(C+N)/Fe] and
[C/N] arising from these sources for two stars, HD~122563 (\Teff= 4600
K) and CS~22948--066 (\Teff= 5100 K), at the cool and warm end of our
sample.

The errors are very similar in both cases and thus are practically 
independent of the temperature of the star between 4600 and 5100K.
Since log $g$ is derived from the ionization equilibrium, it is not
independent of \Teff, so a change in \Teff\ also affects log g.  The
total error budget therefore contains significant covariance terms.
To minimise these effects we have computed [C/Fe], [N/Fe] and 
[(C+N)/Fe] relative to Fe~I for each model, since these abundances are
affected similarly by a change in temperature.

The total error has been approximated by $\sqrt(\sigma_{m}^{2} +
\sigma_{T}^{2} + \sigma_{g}^{2} + \sigma_{v}^{2})$ where $\sigma_{m}$
is the measurement error (Table \ref {tabund}) and $\sigma_{T}$,
$\sigma_{g}$, $\sigma_{v}$ are derived from Table \ref {errorsN}
($\Delta (D-A)$, $\Delta (B-A)$, $\Delta (C-A)$). Note that in 
Table \ref{errorsN} the assumed change in $T_{eff}$ is 100 K.

% TABLE 2
\begin {table}[t]
\caption {Changes ($\Delta$) in the derived N abundance caused by errors in 
model atmosphere parameters for HD~122563, a cool star, and CS~22948-066 
(hotter). A is the adopted model; B, C, and D vary log $g$, \vt, and \Teff\ 
individually as shown, while in model E, log $g$ and \vt\ have been re-adjusted 
for consistency with the lower \Teff (See also Paper V, Tables 6 and 7).}
\label {errorsN}
\begin {center}
\begin {tabular}{cccccc}
\hline \hline
\multicolumn {3}{l}{HD~122563}\\
\multicolumn {5}{c}{A: T$_{eff}$= 4600 K, log g= 1.0 dex, vt= 2.0 
km s$^{-1}$}\\
\multicolumn {5}{c}{B: T$_{eff}$= 4600 K, log g= 0.9 dex, vt= 2.0 
km s$^{-1}$}\\
\multicolumn {5}{c}{C: T$_{eff}$= 4600 K, log g= 1.0 dex, vt= 1.8 
km s$^{-1}$}\\
\multicolumn {5}{c}{D: T$_{eff}$= 4500 K, log g= 1.0 dex, vt= 2.0 
km s$^{-1}$}\\
\multicolumn {5}{c}{E: T$_{eff}$= 4500 K, log g= 0.6 dex, vt= 1.8 
km s$^{-1}$}\\
\hline
~   &$\Delta_{B-A} $ & $\Delta_{C-A} $& $\Delta_{D-A} $& $\Delta_{E-A} $\\
$[Fe/H]$      &  0.03 &  0.03 & -0.11 &  0.03\\
~\\
logN(Li)      &  0.01 &  0.00 & -0.12 & -0.06\\
~\\
logN(C)       &  0.04 &  0.00 & -0.23 & -0.08\\
$[C/Fe~I]$    &  0.02 & -0.09 & -0.03 & -0.05\\
~\\
logN(N)       &  0.05 &  0.00 & -0.30 & -0.05\\
$[N/Fe~I]$    &  0.05 & -0.13 & -0.14 & -0.06\\
~\\
$[C/N]$       & -0.01 &  0.00 &  0.10 & 0.00\\
~\\
$[(C+N)/Fe~I]$&  0.07 & -0.02 & -0.24 & 0.05\\
\hline \hline
~\\
\multicolumn {3}{l}{CS~22948--066}\\
\multicolumn {5}{c}{A: T$_{eff}$= 5100 K, log g= 1.8 dex, vt= 2.0 
km s$^{-1}$}\\
\multicolumn {5}{c}{B: T$_{eff}$= 5100 K, log g= 1.7 dex, vt= 2.0 
km s$^{-1}$}\\
\multicolumn {5}{c}{C: T$_{eff}$= 5100 K, log g= 1.8 dex, vt= 1.8 
km s$^{-1}$}\\
\multicolumn {5}{c}{D: T$_{eff}$= 5000 K, log g= 1.8 dex, vt= 2.0 
km s$^{-1}$}\\
\multicolumn {5}{c}{E: T$_{eff}$= 5000 K, log g= 1.5 dex, vt= 2.0 
km s$^{-1}$}\\
\hline 

~&$\Delta_{B-A} $ & $\Delta_{C-A} $& $\Delta_{D-A} $& $\Delta_{E-A} $\\
$[Fe/H]$    & -0.02 & +0.02 & -0.05 & -0.11\\
~\\
logN(Li)    &  0.01 &  0.00 & -0.11 & -0.10\\
~\\
logN(C)     & +0.04 &  0.00 & -0.20 & -0.10\\
$[C/Fe~I]$    & +0.04 & -0.05 & -0.09 & +0.00\\
~\\
logN(N)     & +0.05 &  0.00 & -0.30 & -0.10\\
$[N/Fe~I]$   & +0.05 & -0.05 & -0.19 & -0.01\\
~\\
$[C/N] $    & -0.01 &  0.00 &  0.07 & -0.03\\
~\\
$[(C+N)/Fe~I]$&  0.07 & -0.02 & -0.24 &  0.01\\
\hline 
\end {tabular}
\end {center}
\end {table}

\section {Mixing in metal-poor giants}

\subsection {Evidence from Carbon and Nitrogen} \label{BehavCN}

The aim of this paper is to study the synthesis of the light elements in the 
Galaxy at the earliest times, i.e. their abundances in gas that was enriched by 
the first Type II supernovae. Generally, the convective atmosphere of a cool 
star is a good tracer of the chemical composition of the interstellar gas at the 
time and place of its formation. However, in giant stars material from deeper 
layers may be dredged to the surface and thereby alter the  initial 
composition (e. g. Gratton et al. \cite{GSC00}), so we must carefully check if such 
mixing has occurred in our stars. 

In dredged-up matter that has been processed by the CNO cycle, N has been 
produced at the expense of C, and the atmosphere of the star becomes ``N rich'' 
and ``C poor'' relative to its initial composition (thus, mixing during any 
first dredge-up episodes cannot explain the C-rich stars, especially in our 
low-mass stars.)

%FIG 6
\begin {figure} 
\begin {center}
\resizebox  {6.cm}{6.0cm} 
{\includegraphics {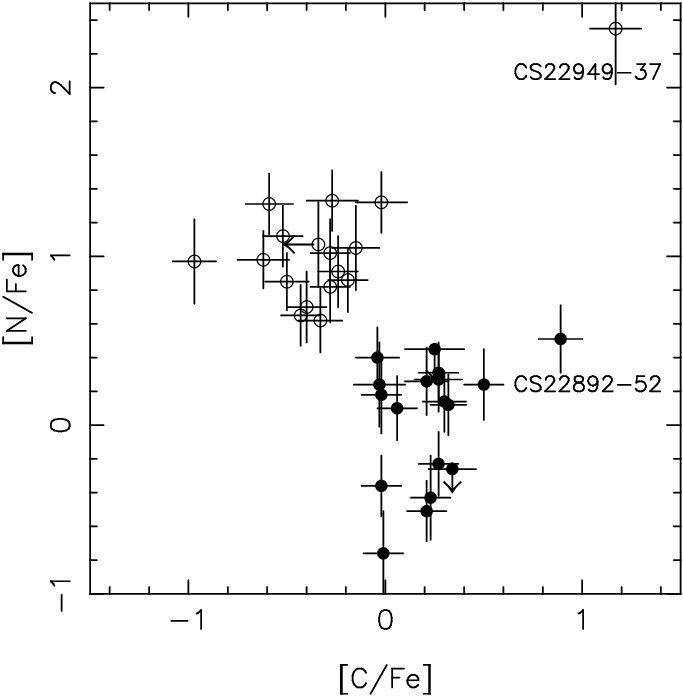} }
\caption {[N/Fe] vs.  [C/Fe] for our sample.  Two groups are clearly
separated: the ``mixed'' stars ([N/Fe]$>0.5$) shown as open circles, and
the ``unmixed'' stars ([N/Fe]$<0.5$) shown as dots (both with error
bars).  CS~22892-52 and CS~22949-37 are the two peculiar carbon-rich
stars.  }
\label {cnfe}
\end {center}
\end {figure}

Fig. \ref{cnfe} shows [N/Fe] vs. [C/Fe] for all stars in our sample. The stars 
clearly fall in two separate groups: the apparently unmixed stars with 
[C/Fe]~$\ge 0.0$ and [N/Fe]~$< +0.5$, and stars with [C/Fe]~$< 0.0$ and 
[N/Fe]~$> +0.5$ which show evidence of mixing. These ratios are, by themselves, 
no proof that the stars of the second group have been mixed internally; the gas 
from which they were formed {\it could} have had this composition. However, for 
simplicity we will continue to refer to these two groups as the unmixed and 
mixed stars, respectively. The 'mixed' stars are identified by an ``m'' in Table 
\ref {tabund} (last column).

\subsection {Evidence from the C/N ratio}

\begin {figure}
\begin {center}
\resizebox  {8.cm}{5.0cm} 
{\includegraphics {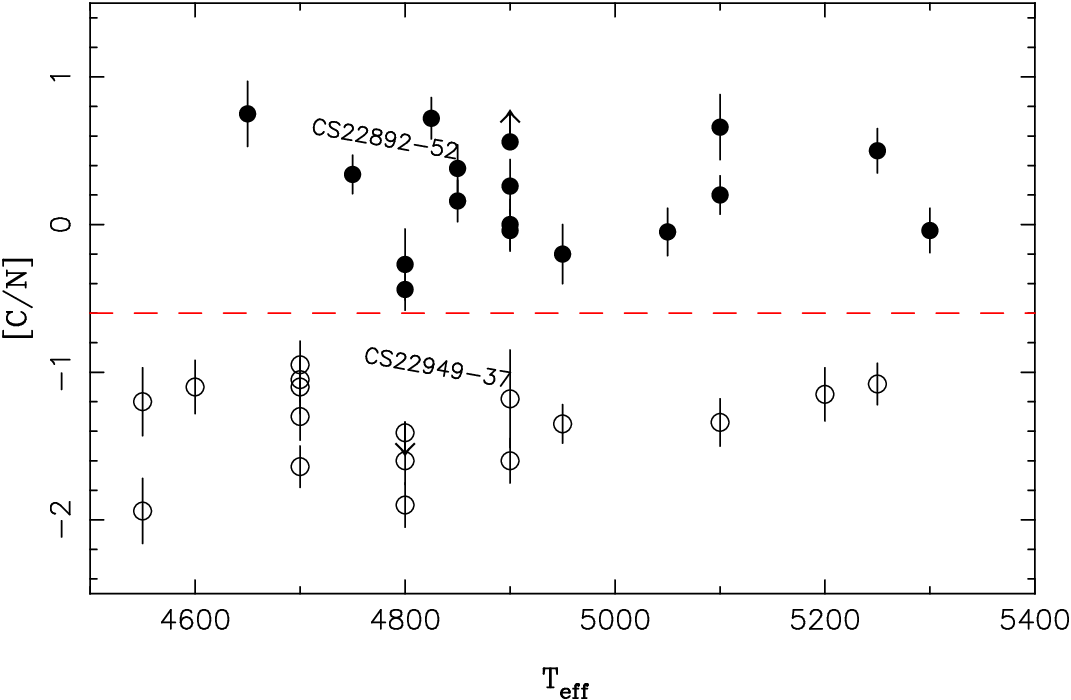} }
\caption {[C/N] vs. \Teff~ for the sample; symbols as in Fig. \ref{cnfe}. The 
line [C/N]= --0.4 separates the mixed and unmixed stars. Stars with only upper 
limits on C and/or N abundance have been omitted. Note that most unmixed stars 
have \Teff $\geq$ 4800 K (cf. also Paper~V, sect. 4.1).}
\label {cnteff}
\end {center}
\end {figure}

Because the CNO process turns C into N, the [C/N] ratio is a sensitive indicator 
of mixing. Fig. \ref{cnteff} shows the [C/N] ratio vs. \Teff. Like in Fig. 
\ref{cnfe}, the mixed and unmixed stars are well separated.  Note that most of 
the unmixed stars have \Teff $\geq$ 4800 K (cf. Paper~V, Section 4.1).

%FIG 7

Two stars in our sample are known to be C rich: CS~22892--052 (Sneden et al. 
\cite{SCL03}) falls close to the unmixed stars in Fig. \ref{cnfe}, while 
CS~22949--037 (Depagne et al. \cite{DHS02}) is quite peculiar, as its huge N 
abundance is suggestive of mixing (consistent with its low ${\rm ^{12}C/^{13}C}$ 
ratio), yet it falls quite far from the other mixed stars. None of these two 
stars appears unusual in Fig. \ref{cnteff}, however.

For three stars we have only upper abundance limits: CD--38:245 (C only), 
BS~16477-003 (N only), and BS~16467--032 (both C and N); they are marked with 
arrows in Fig. \ref{cnfe}, but omitted in Fig. \ref{cnteff} as the [C/N] ratio 
is undefined in these stars.

\subsection {Evidence from Lithium}

When low-mass stars, such as those in our sample, evolve through the subgiant 
and red giant branches, the surface convection zone progressively deepens. This 
mixes the stellar atmosphere with material from deeper layers in which Li has 
been depleted by nuclear burning, and reduces the observed Li abundance of the 
star. The degree of dilution increases as the convective zone penetrates deeper; 
thus, one expects a steady decline of the observed Li abundance as a star 
ascends the RGB (Pilachowski et al., \cite {PSB93}). If mixing of the surface 
material is deep enough to reach the layers where C is burned into N, the 
atmospheric Li will {\em a fortiori} burn away rapidly. The mixed stars 
should thus show 
much lower Li abundances than the unmixed stars. We have performed this test.

\begin {figure}
\begin {center}
\resizebox  {7.5cm}{5.0cm} 
{\includegraphics {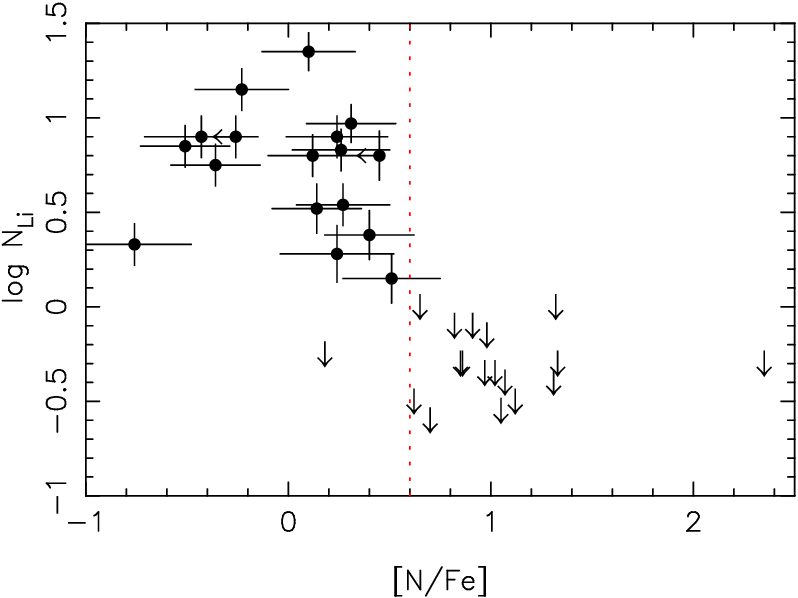} }
\resizebox  {8.cm}{0.5cm} 
{\includegraphics {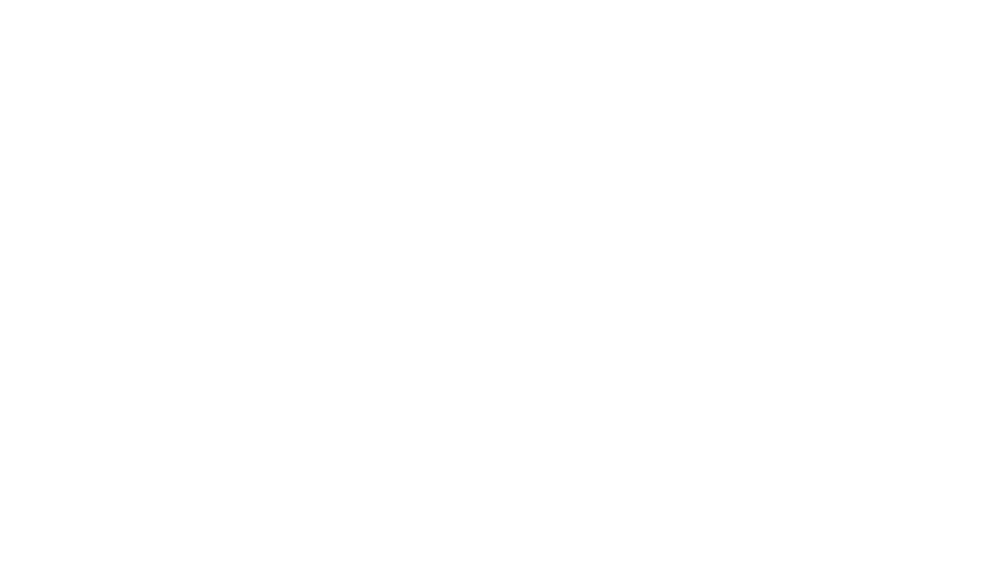} }
\resizebox  {7.5cm}{5.0cm} 
{\includegraphics {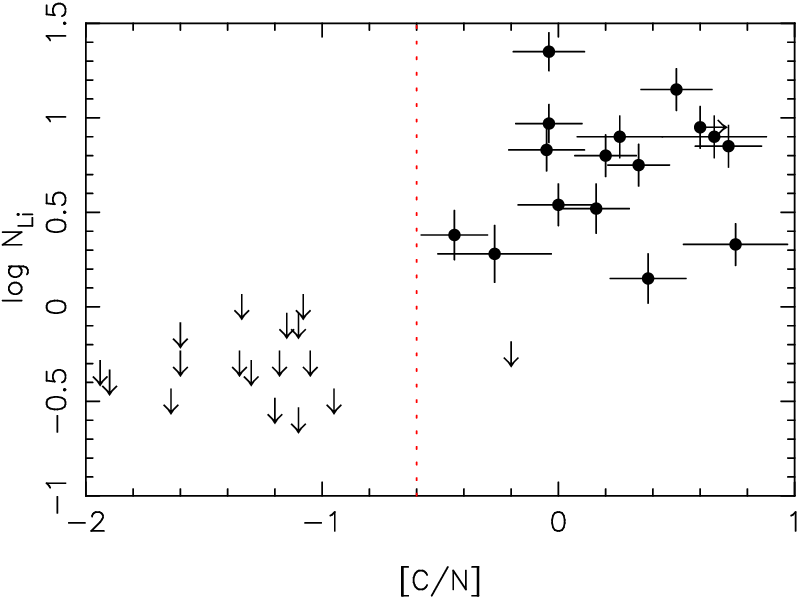} }
\caption {
Lithium abundances vs. [N/Fe] and [C/N] (symbols as in Fig. \ref{cnfe}). All the 
mixed stars ([N/Fe] $>0.5$ or [C/N] $<-0.5$; dotted lines) have destroyed their 
original Li.} 
\label {nli}
\end {center}
\end {figure}

Fig. \ref {nli} shows the observed surface abundance of Li as a function of
[N/Fe] and [C/N].  The Li line is not detected in any of the mixed stars 
([N/Fe]$>0.5$ or [C/N]$<-0.5$), with an upper limit to the Li abundance of log 
N(Li)$<$0. In contrast, we find log N(Li)$> 0.15$ for all the unmixed stars, 
with the sole exception of CS~30325--094 (log N(Li) $< -0.25$). Several effects 
may deplete the fragile Li in individual stars (rotation, gravity waves, binary 
interaction, \ldots) so we have not sought to clarify this single anomaly.

%FIG 8

The consistent results from the [N/Fe] and [C/N] ratios and the Li abundance are 
strong evidence that the stars with [N/Fe]$>0.5$ or [C/N]$<-0.5$ did experience 
mixing between deeper CNO-processed layers and the surface of the stars.

\subsection {Location in the H-R diagram}

Changes of abundances have been detected as stars evolve along the RGB in both 
globular cluster giants (Kraft \cite{Kr94}) and field stars (Gratton et al. 
\cite{GSC00}). It would thus be interesting to study the positions of our XMP 
stars in an H-R diagram to see at which phase of the evolution mixing has 
occurred. Unfortunately, these stars are too distant to have usable parallaxes, 
so their precise distances and luminosities are unknown.

%FIG 9
\begin {figure}
\begin {center}
\resizebox  {5.cm}{8.0cm} 
{\includegraphics {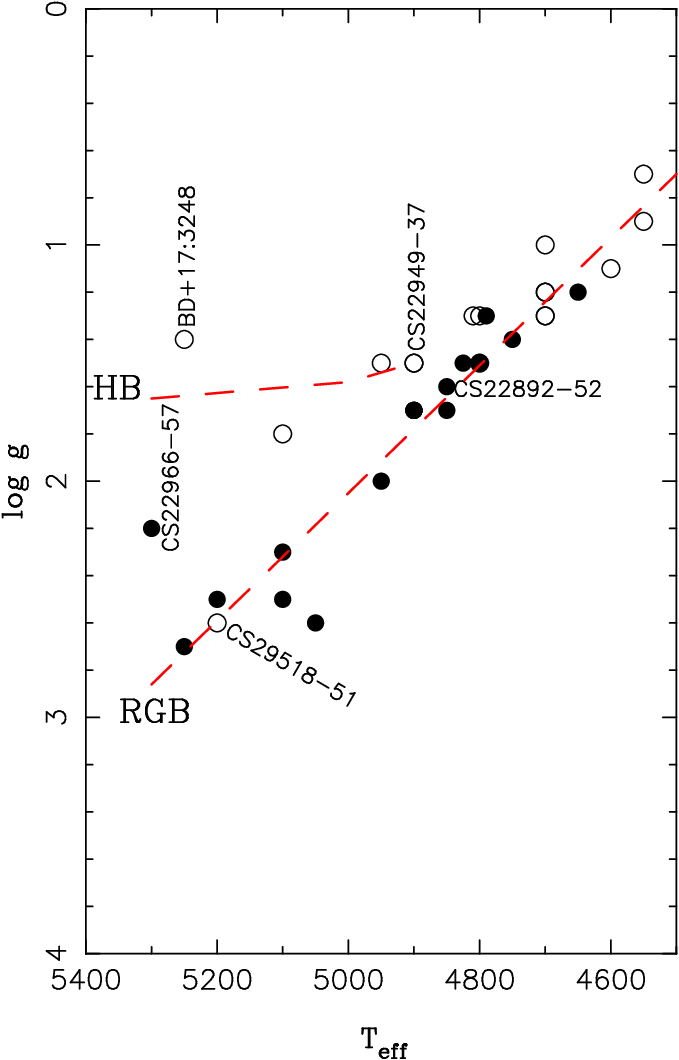} }
\caption {
log $g$ vs. log \Teff\ diagram for the sample (symbols as in Fig. \ref{cnfe}). 
The RGB and HB are fairly well-defined (dashed lines). At fixed \Teff, the 
unmixed stars on the ``low'' RGB have higher log $g$ than the mixed stars.}
\label {hr}
\end {center}
\end {figure}

However, the spectroscopic value of log $g$ is a first-order indicator
of the luminosity of the star.  These values of log $g$ may suffer
from NLTE effects and thus be different from the true gravity, but
Gratton et al.  (\cite{GSC00}) argue that NLTE corrections to log $g$
are smaller than suggested in most previous literature; moreover, our
stars lie in relatively narrow intervals of temperature and gravity,
and NLTE effects on iron should be constant in the metallicity range
of our stars (Korn, Shi, \& Gehren, \cite {KSG03}).
Therefore, any corrections would be similar for all the stars, and 
abundance trends with log $g$ should be robust against NLTE effects. 

Fig. \ref{hr} shows the log $g$ - log \Teff~ diagram for our sample. The unmixed 
stars form a fairly well-defined lower RGB, while virtually all stars on and 
above the HB have the spectroscopic characteristics of mixed stars, as 
previously found for moderately metal-poor field stars by Gratton et al.  
(\cite{GSC00}). Three stars deviate from the general trend and are discussed in 
the following: 

BD~17:3248, a mixed star strongly enriched in $r$-process elements (Cowan et al. 
\cite{CSB02}), has been classified as an RHB star (Bond \cite{Bon80}, 
Pilachowski et al. \cite{PSK96}, Alonso et al. \cite{AAM98}), in good agreement 
with its position in Fig. \ref{hr}. In contrast, the Hipparcos parallax, $\pi = 
3.67\pm1.5$mas, would place BD~17:3248 on the subgiant branch. However, Cowan et 
al. (\cite{CSB02}) conclude, from Str\"omgren photometry, that the star is 
indeed highly evolved and the small Hipparcos parallax probably unreliable. 

CS~22966-057, the hottest star of our sample, falls between the lower RGB and 
the HB. Our two spectra of this star from September and October 2001 yield 
radial velocities of +100.7 km s$^{-1}$ and +103.0 km s$^{-1}$, a variation too 
large to be explained by our observational error, 0.5 km s$^{-1}$. This star is 
probably a binary.

CS~29518--051 falls on the low RGB, but has all the characteristics of a mixed 
star. A large error in its gravity is unlikely. Our single spectrum 
gives no information on any velocity variations, but the star clearly deserves 
close attention in the future.

\subsection {Deep mixing and the early Galactic O abundance }

\begin {figure}
\begin {center}
\resizebox  {8.0cm}{4.75cm} 
{\includegraphics {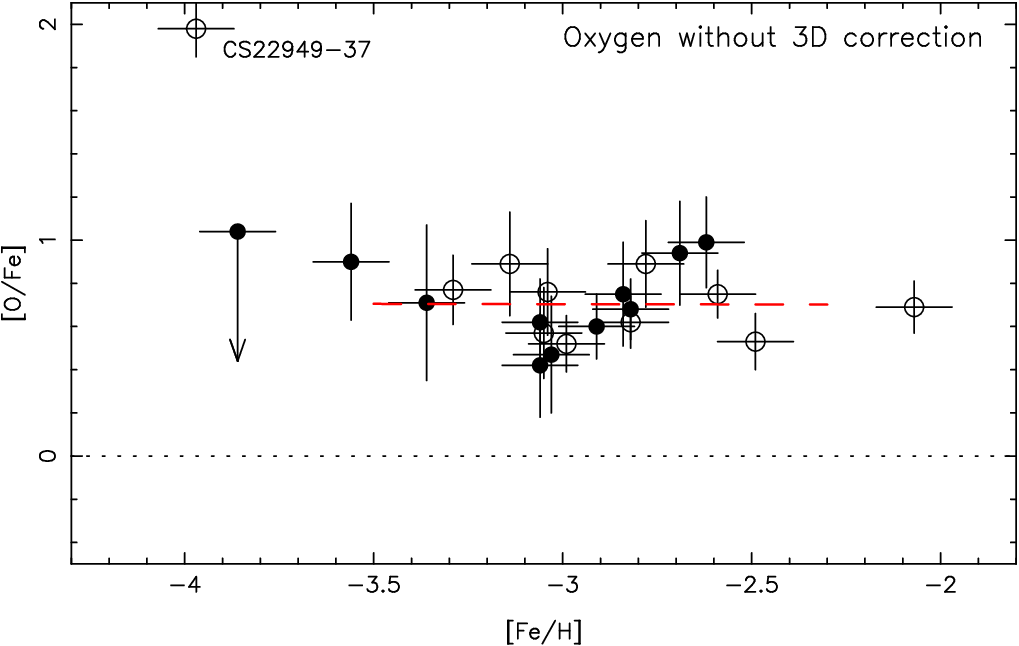} }
\caption {
[O/Fe] vs.  [Fe/H] for the sample; symbols as in Fig.  \ref{cnfe}.
The lack of any significant difference between mixed and unmixed stars
is strong evidence that deep mixing has not occurred (the O abundances
shown here are not corrected for 3D effects; cf.  Paper V).}
\label {oxy-mix}
\end {center}
\end {figure}

A very deep mixing event might bring matter to the surface in which the O-N 
cycle has partially transformed O into N. It is thus interesting to compare the 
O abundances of the mixed and unmixed stars. As seen in Fig. \ref{oxy-mix}, 
there is no significant difference in [O/Fe] between mixed and unmixed stars, 
and thus no evidence for contamination by deep mixing, as might occur in an AGB 
star of an earlier generation or in a hypothetical binary companion. This result 
is strong evidence that the O abundances determined in Paper V are indeed 
reliable indicators of the very early nucleosynthesis in the Galaxy.

%FIG 10

\subsection {CNO abundance ratios as detailed diagnostics of mixing}

\begin {figure}
\begin {center}
\resizebox  {8.0cm}{4.75cm} 
{\includegraphics {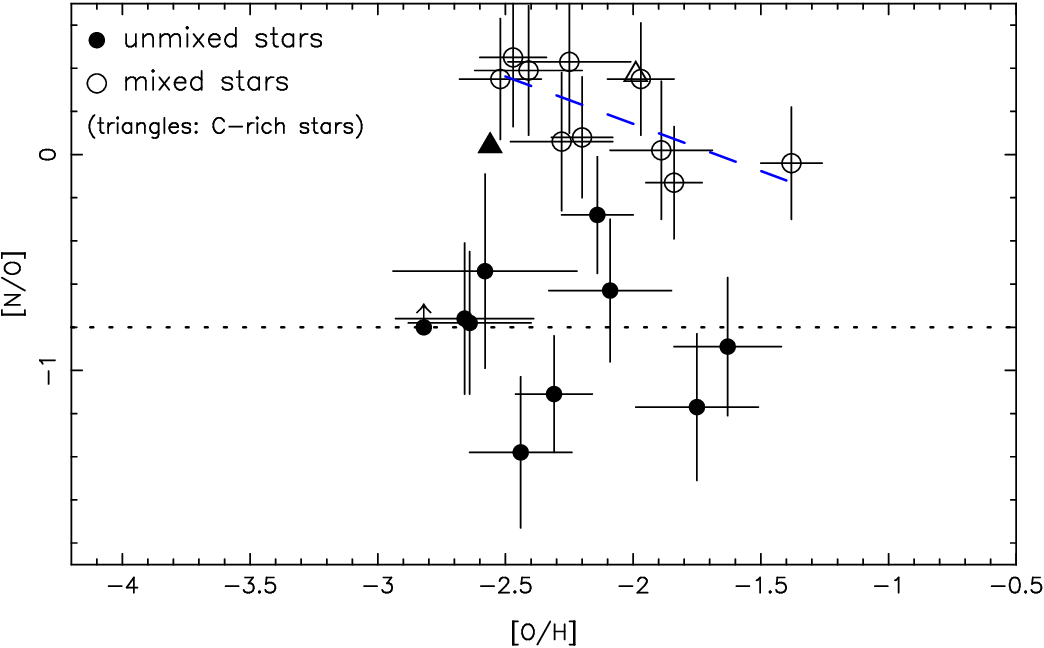} }
\caption {
[N/O] vs.  [O/H] for the sample; symbols as in Fig.~\ref{cnfe} (The
carbon stars CS~22892--052 and CS22949--037 are shown as triangles).
The [N/O] vs.  [O/H] relation in the mixed stars is quite tight, [N/O]
decreasing slightly with increasing [O/H] (dashed line).  This
suggests that the amount of N mixed to the surface does not depend
strongly on metallicity.
}
\label  {NsOmix} 
\end {center}
\end {figure}

Fig.  \ref {NsOmix}  compares the [N/O] ratios in mixed and unmixed
stars.  As expected, [N/O] is systematically higher in the mixed
stars, which have themselves produced N from C in the CN cycle.  This
additional N must therefore be of secondary origin\footnote{Note that
a high N abundance in a metal-poor star does not necessarily imply
that its atmosphere has been mixed with its CN processing interior.  A
few unmixed, N-rich, metal-poor stars are known to exist (e.g. G64-12,
Israelian et al.  2004).}.  Moreover, [N/O] and [O/H]
are tightly correlated in all the mixed stars, in contrast with the
large scatter observed in the unmixed stars.  The large amount of N
brought to the surface of the mixed stars has thus erased the large 
scatter of the initial N abundances.  If this amount of N
brought to the surface 
is assumed to be roughly independent of the metallicity of the star,
the [N/O] ratio should decrease with increasing [O/H], as indeed seen
in the data  (Fig. \ref {NsOmix}, mixed stars, dashed line).

%FIG 11

%FIG 12
\begin {figure}
\begin {center}
\resizebox  {8.0cm}{4.75cm} 
{\includegraphics {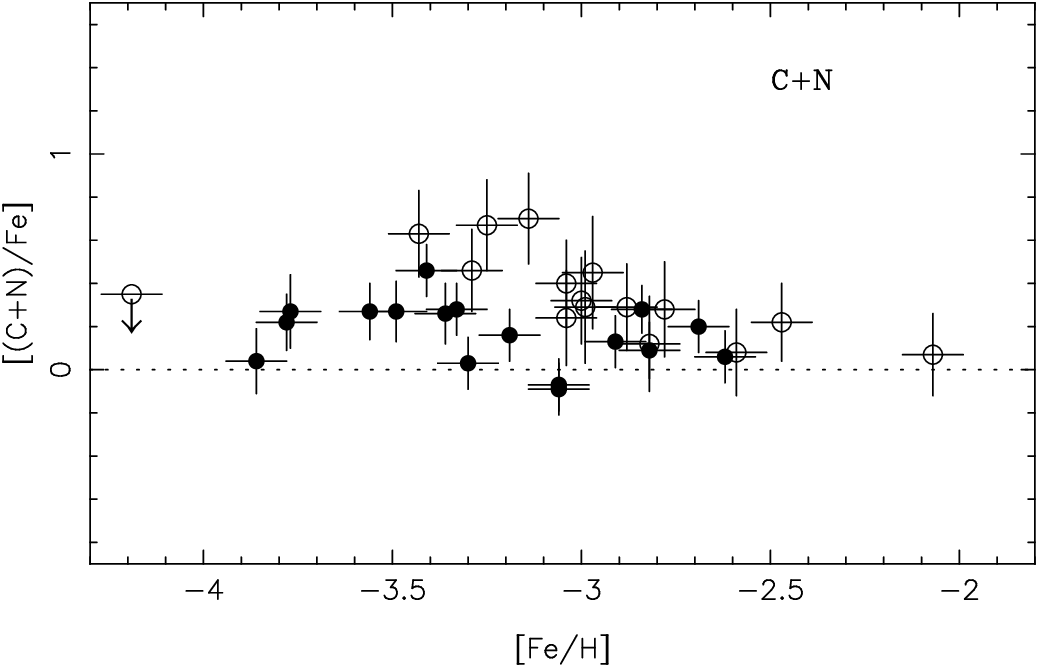} }
\caption {[(C+N)/Fe] vs.  [Fe/H]; symbols as in Fig.  \ref{cnfe}.  The
C+N abundance shows much smaller scatter than C or N alone (cf.  Figs.
\ref{Carbrut} and \ref{Azbrut}).  $<$[(C+N)/Fe]$>$ $\approx0.25$ dex
at low [Fe/H].  CS~22949--037 and CS~22892--052 are not shown here.}
\label{C+N}
\end {center}
\end {figure}

Because we find no evidence for processing by the O-N cycle in the
mixed stars, if the excess of N is due to internal mixing it results
only from the transformation of C nuclei into N.  Thus we would expect
that the mean value of the C+N abundance in the mixed and unmixed 
stars should be the same. Fig.~\ref{C+N} shows that within errors this 
hypothesis is compatible with the observations. Note that the three
mixed stars which stand out towards the high [(C+N)/Fe] ratios 
may belong to the horizontal branch.

%BUT C+N SEEMS TO HAVE A SLOPE IN THE MIXED STARS ?? / JA
%Yes BUT WHY ???? Monique

The peculiar C-rich stars CS~22949--037 and CS~22892--052 cannot be compared
directly to the other stars of the sample. Both are included in Fig.
\ref {NsOmix}, where CS~22892--052 has a rather high [N/O] ratio for an unmixed 
star while CS~22949-037 falls among the other mixed stars. Both have been 
omitted from Fig. \ref{C+N}, where they would fall far from the other stars in 
their groups.\\
~\\
~\\
%\subsection {Existence of internal mixing in some extremely metal-poor 
%giants}

The consistent evidence discussed above shows that our sample divides cleanly 
into two groups. In one, the stars have experienced mixing of their atmospheres 
over their lifetimes; in the other they have not.

\section {CNO elements in the early Galaxy}

 We have 
presented strong evidence that the atmospheric CNO abundances of the 
unmixed stars represent the original abundances of these species in the gas from 
which the stars formed. We discuss the consequences of this identification in 
the following.

\subsection {Carbon and the [C/O] ratio}

%FIG 13
\begin {figure}
\begin {center}
\resizebox  {8.0cm}{4.75cm} 
{\includegraphics {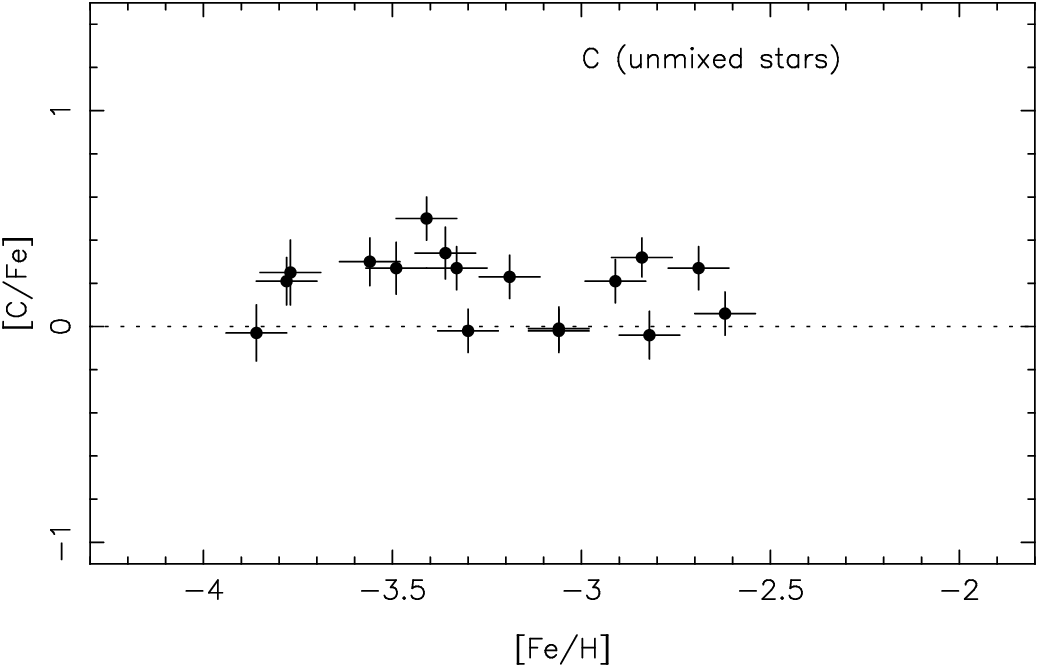} }
\resizebox  {8.0cm}{0.5cm} 
{\includegraphics {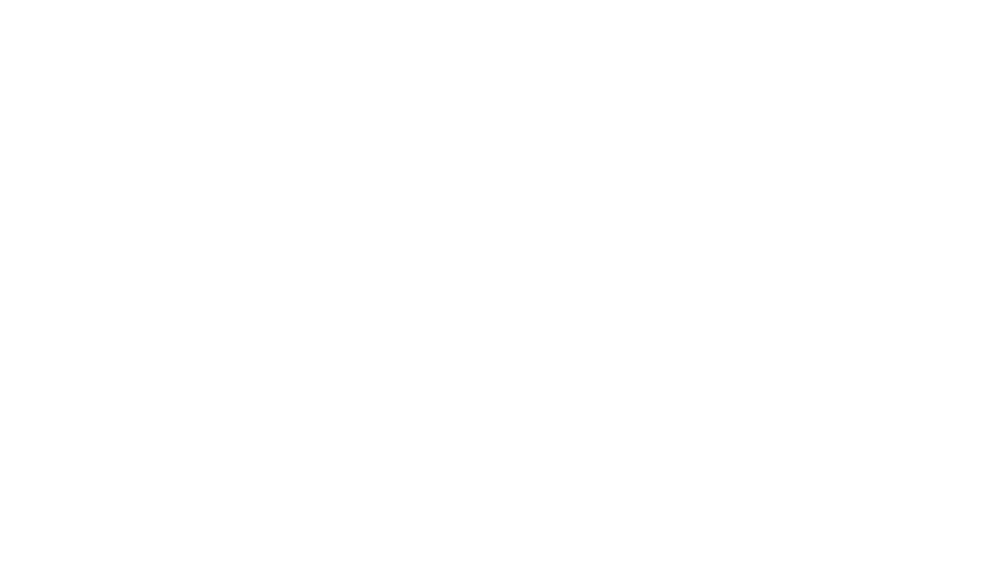} }
\resizebox  {8.0cm}{4.75cm} 
{\includegraphics {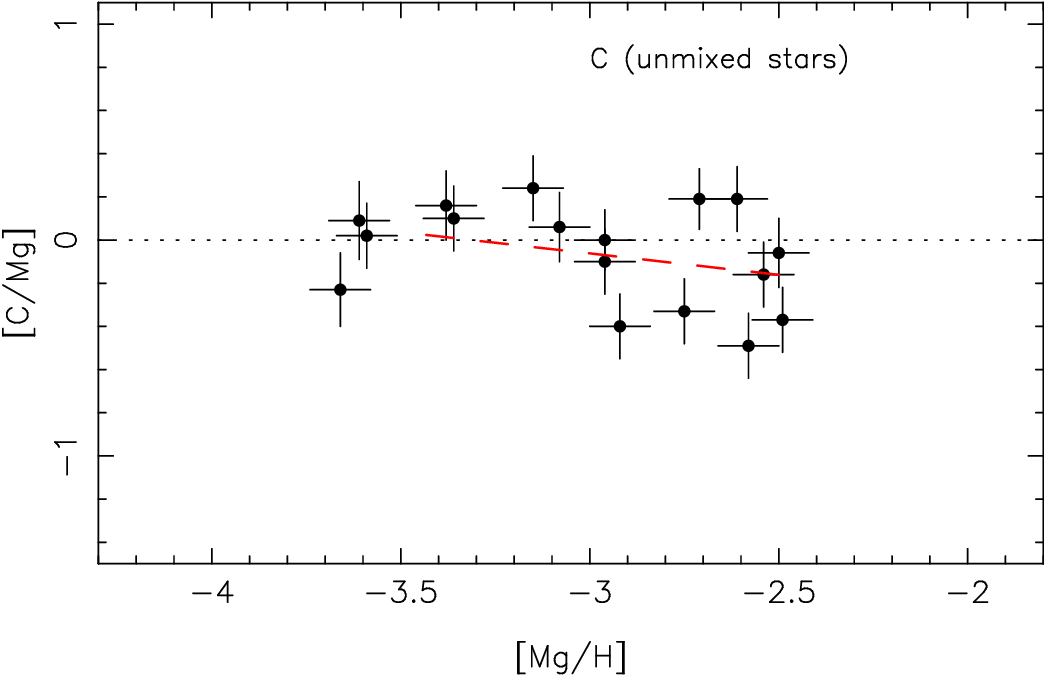} }
\caption {[C/Fe] and [C/Mg] vs. [Fe/H] and [Mg/H] for the unmixed stars. [C/Mg] 
may increase slightly towards lower metallicity; for [Mg/H] $< -2.9$, 
$<$[C/Mg]$>\approx$ 0.0 dex.}
\label {C-C-notmix}
\end {center}
\end {figure}

%FIG 14
\begin {figure}
\begin {center}
\resizebox  {8.0cm}{4.75cm} 
{\includegraphics {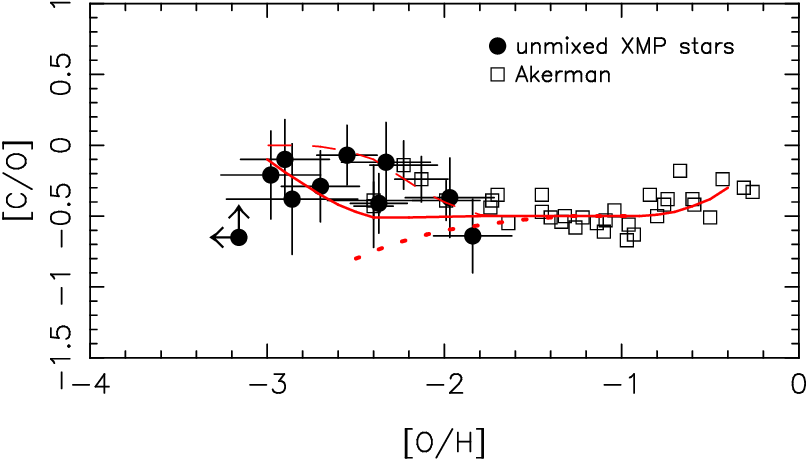} }
\caption {[C/O] vs.  [O/H] for our unmixed stars (dots) and the halo
stars from Akerman et al.  (\cite{ACN04}; squares).  The lines show
the predictions of their standard model for C and O, using the yields
of Meynet \& Maeder (\cite{MM02}, dotted), Chieffi \& Limongi
(\cite{CL02}, dashed), and Chieffi \& Limongi with a top-heavy IMF ($M
> 10M_{\odot}$, solid line).  The data seem to favour substantial
early C production in massive zero-metal supernovae.  }
\label {COAker}
\end {center}
\end {figure}

Fig.~\ref{C-C-notmix} shows the C abundance in our unmixed stars, with
both Fe and Mg as reference elements.  The [C/Fe] vs.  [Fe/H] relation
is remarkably flat; [C/Fe] $\approx +0.18\pm0.16$ dex over the entire
range ${\rm -4.0 < [Fe/H] < - 2.5}$.  The dispersion of the [C/Mg]
ratios is only slightly larger than for [C/Fe]; a mild decrease of
[C/Mg] with increasing metallicity is suggested.

Akerman et al.  (\cite{ACN04}) recently studied the variation of the
[C/O] ratio vs.  [O/H] in metal-poor halo stars.  At low metallicity
($\rm{[O/H] < -2.0}$), [C/O] is expected to decrease because O from
massive supernovae increasingly dominates the enrichment process,
while C is produced in stars of all masses.  However, Akerman et al.
(\cite{ACN04}) found no decrease in [C/O] at low metallicity; on the
contrary, [C/O] appeared to actually increase below [O/H]= --1.5,
possibly to near-solar values at the lowest metallicities.

Fig.  \ref{COAker} shows [C/O] vs.  [O/H] for the combined data set;
our stars extend to much lower metallicities than those of Akerman et
al.  Although the O and C indicators used in these two studies are
completely different (high-excitation C and O lines in Akerman et
al., CH and [OI] lines in our case), the results agree nicely where
the samples overlap.  Our XMP stars confirm that [C/O] increases to
near solar values at very low metallicity.

Fig. \ref{COAker} also shows the predictions of different chemical evolution 
models, following Akerman et al. (\cite{ACN04}):
\begin{itemize} 
\item Their standard model using the Meynet \& Maeder (\cite{MM02}) yields for 
massive stars ($8\leq M\leq 80 M_{\odot}$) and the Kroupa et al. (\cite{KTG93}) 
IMF (dotted line).
\item Their standard model with the same IMF, but C and O yields from Chieffi \& 
Limongi (\cite{CL02}) for metallicities $0\leq z\leq10^{-5}$ (dashed line). 
Chieffi \& Limongi (\cite{CL02}) argue that the ejecta of metal-free supernovae 
should be carbon rich, because the high core temperature in these stars during 
helium burning would favour the reaction $\rm{3^4He(2\alpha,\gamma)^{12}C}$ over 
$\rm{^{12}C(\alpha,\gamma)^{16}O}$.
\item The same model and Chieffi \& Limongi yields, but with a top-heavy IMF ($M 
\geq 10M_{\odot}$; solid line). 
\end{itemize}
Our new data suggest that the best model would be somewhere between the last two 
of these.

%FIG 15
\begin {figure}
\begin {center}
\resizebox  {8.0cm}{4.75cm} 
{\includegraphics {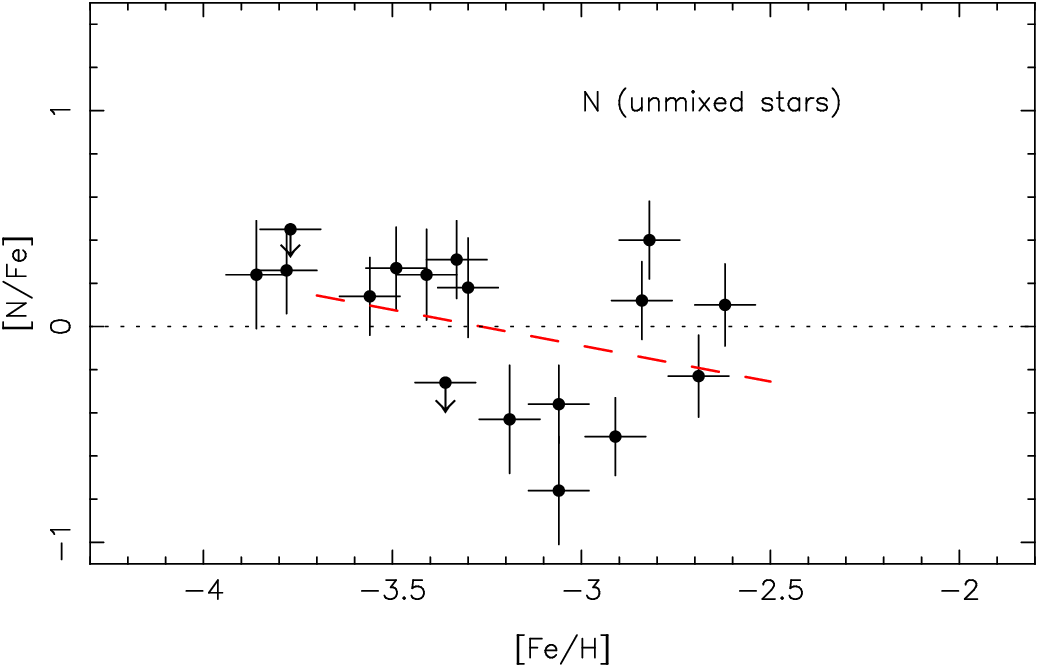} }
\resizebox  {8.0cm}{0.5cm} 
{\includegraphics {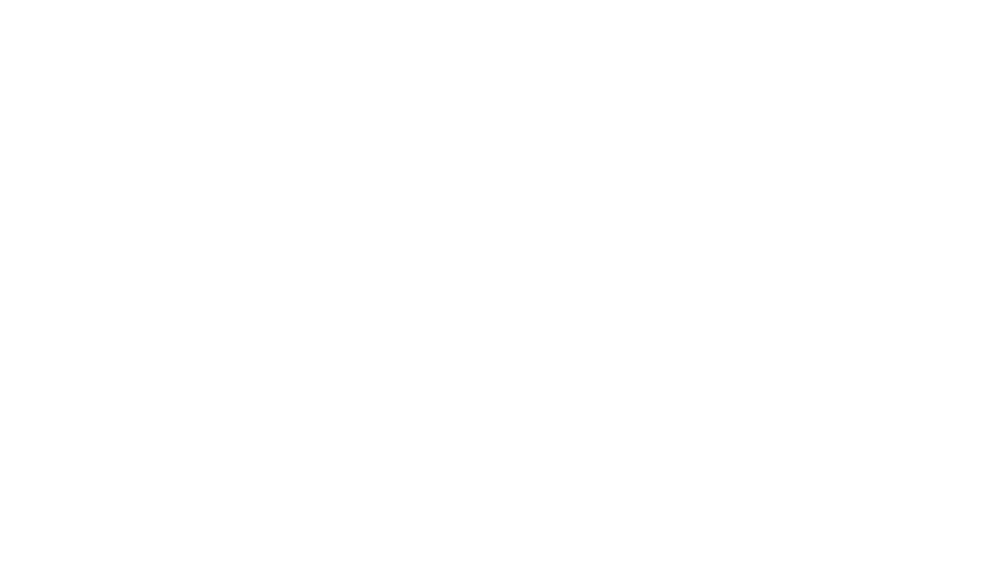} }
\resizebox  {8.0cm}{4.75cm} 
{\includegraphics {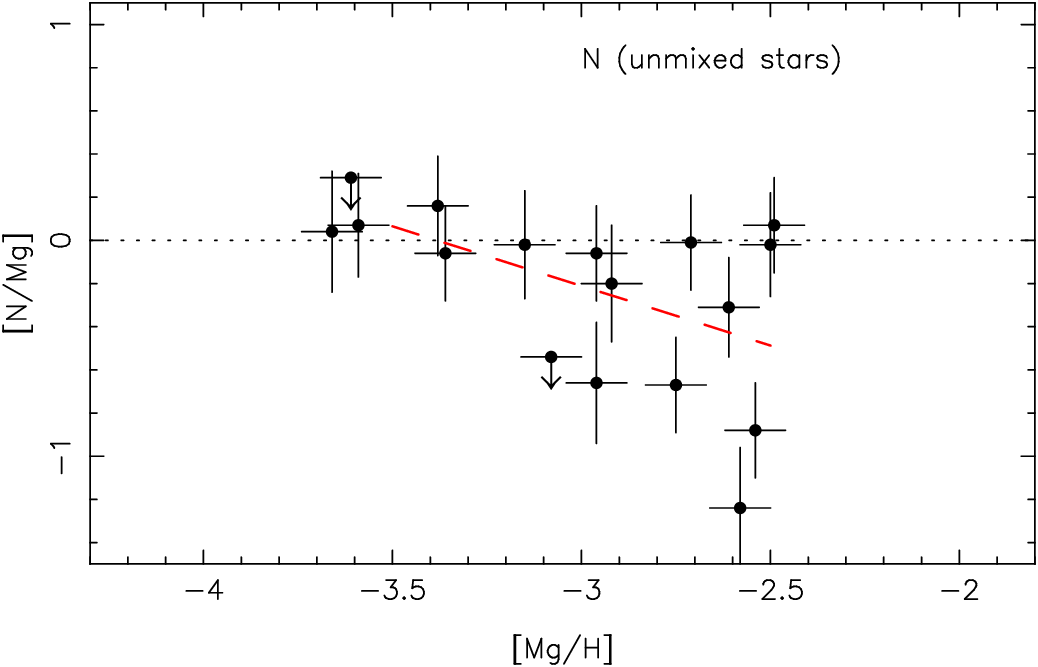} }
\caption {[N/Fe] and [N/Mg] vs. [Fe/H] and [Mg/H] for our unmixed stars. With 
increasing metallicity, [N/Fe] and [N/Mg] decrease (dashed lines) and the 
dispersion increases, a behaviour unexpected from theoretical expectations.}
\label {N-N-notmix}
\end {center}
\end {figure}

\subsection {Nitrogen and the [N/$\alpha$] ratio}

Fig. \ref{N-N-notmix} shows [N/Fe] and [N/Mg] vs. [Fe/H] and [Mg/H] for 
the subsample of unmixed stars. 
The trends of these ratios are not compatible with secondary N production 
in the early Galaxy: instead, a decrease of [N/Mg] and [N/Fe] with 
decreasing metallicity would be expected. The early production of N was 
thus primary. It may be due to massive stars where it would be induced, 
e.g., by mixing between the C producing regions and the H burning 
layer where C is transformed into N. But it could also be due
to contributions by AGB stars.

\begin {figure}
\begin {center}
\resizebox  {8.0cm}{4.75cm} 
{\includegraphics {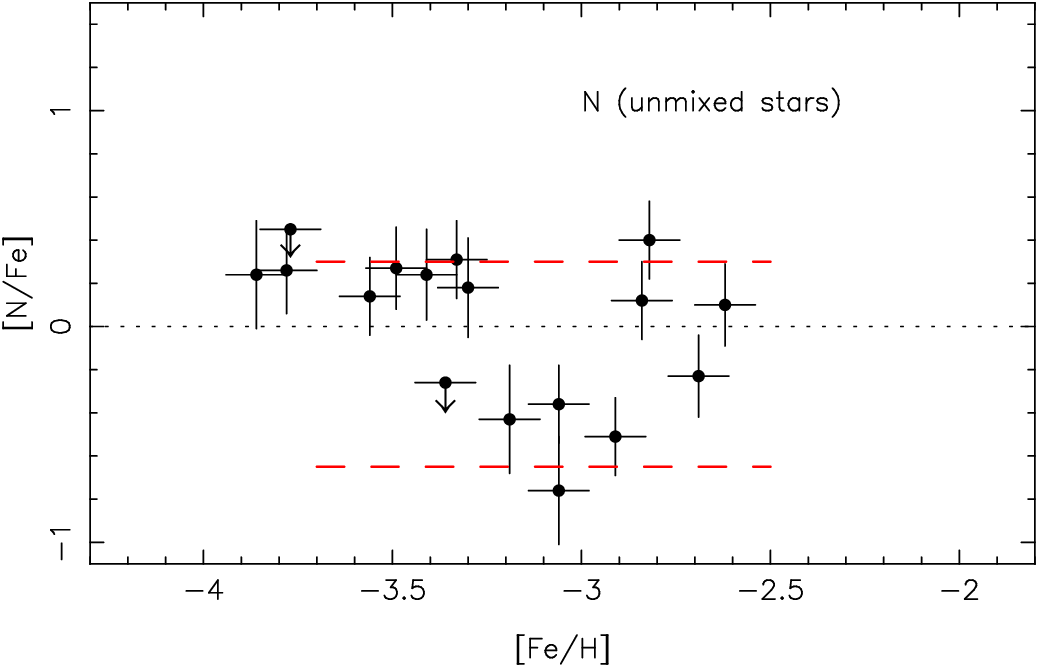} }
\resizebox  {8.0cm}{0.5cm} 
{\includegraphics {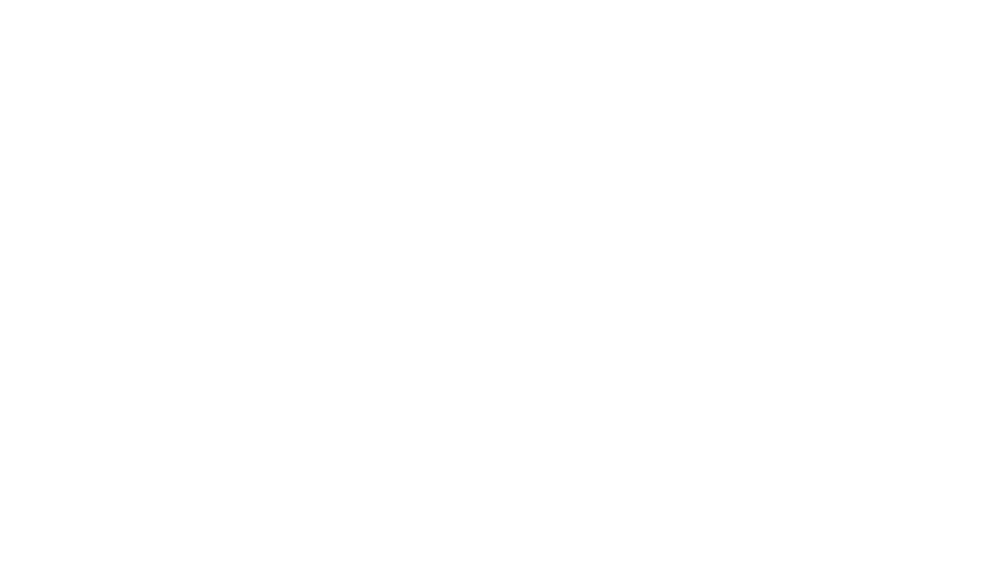} }
\resizebox  {8.0cm}{4.75cm} 
{\includegraphics {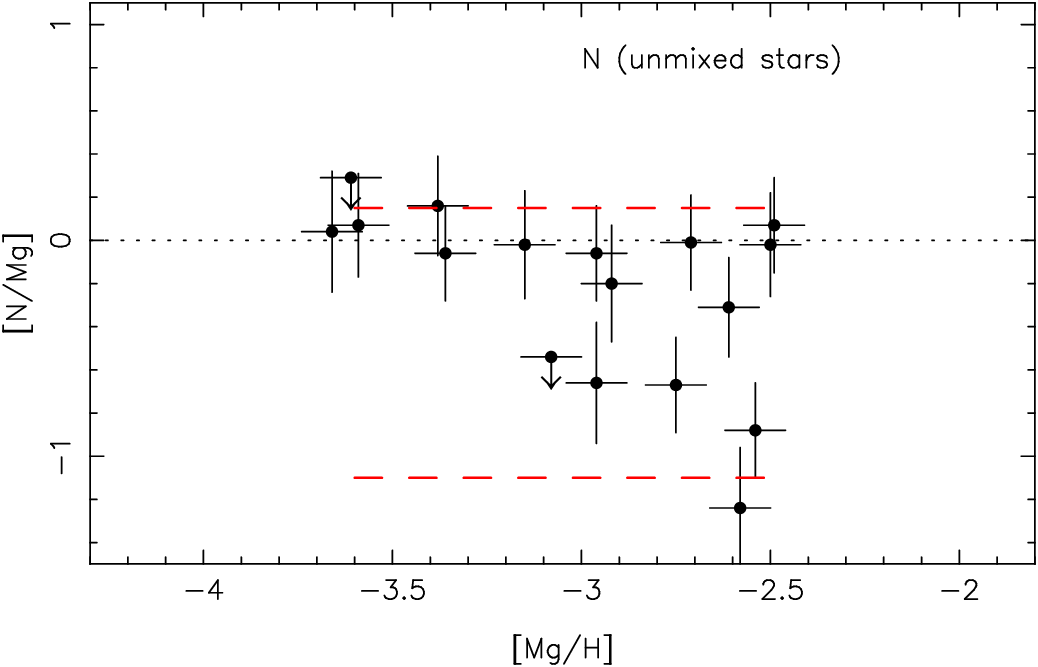} }
\caption {The same data as in Fig. \ref{N-N-notmix}, but interpreted 
differently. We assume here that the low dispersion at low metallicity is a 
spurious effect due to the small size of the sample, while [N/Fe] and 
[N/Mg] in general shows large scatter from star to star, with ${\rm 0.7 < 
[N/Fe]< 0.3}$ and ${\rm -1.1 < [N/Mg]< 0.2}$ as seen for [Fe/H] $>$ --3.3.
}
\label {N-N-notmixb}
\end {center}
\end {figure}

\begin{itemize}
\item    
Fig. \ref{N-N-notmix} appears to show that the mean values of  
[N/Fe] and [N/Mg] decrease with increasing metallicity, perhaps reflecting a 
 decrease in N production relative to both Fe and Mg (the [C/Mg] ratio 
shows a similar trend, but it is much weaker and not significant). The 
scatter in both relations is much larger than for both [C/Fe] and [C/Mg] (see 
Fig. \ref{C-C-notmix}). Similarly to the case of C, the scatter in [N/Mg] 
increases markedly with increasing [Mg/H] from about [Mg/H]= --3.

However, were our N measurements to suffer from {\em
metallicity-dependent} systematic errors, a spurious slope might
result.  3D effects in model atmopheres (see Sec.  \ref{azot1}) might
cause such metallicity-dependent errors. However, although one cannot yet
be certain, it is unlikely that differential effects in stars with [Mg/H] 
between --2.8 and --3.8 could vary the N abundance by the 0.5 dex
needed to eliminate the observed slope (Asplund \cite{Asp04}).
Moreover, such a systematic effect could not explain the increased
dispersion of [N/Fe] and [N/Mg] towards higher metallicity.

{\em Such a decrease of the nitrogen production with increasing
metallicity is not predicted by any massive-star yields at low
metallicities}.\\

\item
Another interpretation is possible, however (Fig.  \ref{N-N-notmixb}):
We have only five N abundance measurements (plus two upper limits) 
in the interval ${\rm -3.7<[Mg/H]<-3}$ where the dispersion seems to
vanish, so the absence of a spread at low metallicity could be just a
statistical effect.  If so, the mean slope drawn in Fig.
\ref{N-N-notmix} would not be real, and [N/Mg] and [N/Fe] would simply
vary from star to star without any mean trend (Fig.
\ref{N-N-notmixb}).  Two possible explanations are:\\
{\bf i)} The maximum [N/Mg] value ($\sim$0.0) reflects the primary
N production by normal massive supernovae, but with some stars showing 
much lower production.  N is not easily produced in massive SNe,
and the yields may depend critically on various
parameters, such as rotation (Maeder \& Meynet \cite{MM02}), explaining
the spread in the observed N abundance.  The spread seems to appear at less
extreme deficiencies (${\rm [Fe/H] > -3.4}$), reflecting the yields of
later, i.e. lower-mass supernovae.\\
{\bf ii)} The lowest N abundances in our stars might represent galactic gas 
enriched by SNe~II, but before any enrichment by massive AGB stars. The majority 
of the N-rich stars in Fig. \ref{N-N-notmixb} would then be formed from matter 
more or less enriched by AGB winds, and the scatter in the N abundances 
would arise from local, inhomogeneous N enrichment  of the gas from which our 
stars formed. In this case, metallicity would not be a good indicator of age. 
Moreover, we would then expect the $s$-process elements to be more 
abundant in N-rich than in N-poor unmixed stars. A preliminary analysis shows 
no clear difference in $s$-process abundances between the two sets of stars;
this will be discussed in more 
detail in a forthcoming paper on the neutron-capture elements in our XMP stars 
(Fran\c cois et al. 2004, in prep.; Paper VII).\\

In both tentative explanations there remains to explain why the
interstellar medium is enriched in N up to a fixed maximum
([N/Fe] $\approx$ 0.3 dex and [N/Mg] $\approx$ 0.2 dex), as clearly seen in
Fig. \ref {N-N-notmixb}.

%WHY DO THE AGB STARS ENRICH THE ISM IN N UP TO A FIXED MAXIMUM? / JA
\end {itemize}

%FIG 16

\subsection {Comparison of halo stars and DLAs}

We have argued that the N abundance in our unmixed XMP stars is very
close to that of the gas from which the stars were formed.  The stars
on the lower RGB have possibly experienced the first dredge-up (if
any), but this should only affect the [N/Fe] or [N/Mg] ratios slightly
(Gratton et al., 2000), even in rapidly rotating stars (Meynet \&
Maeder, 2002).

We now compare our results to the [N/$\alpha$] ratios in DLAs and extragalactic 
H~II regions (or Blue Compact Galaxies, BCGs). E.g., both Prochaska et al. 
(2002) and Centuri\'on et al. (2003) suggest the existence of a plateau at 
[N/$\alpha$] $\sim$ --1.5 dex in addition to the well-known one at [N/$\alpha$] 
= -0.8 dex, which is common to DLAs and BCGs.

%FIG 17
\begin {figure}
\begin {center}
\resizebox  {8.0cm}{4.75cm} 
{\includegraphics {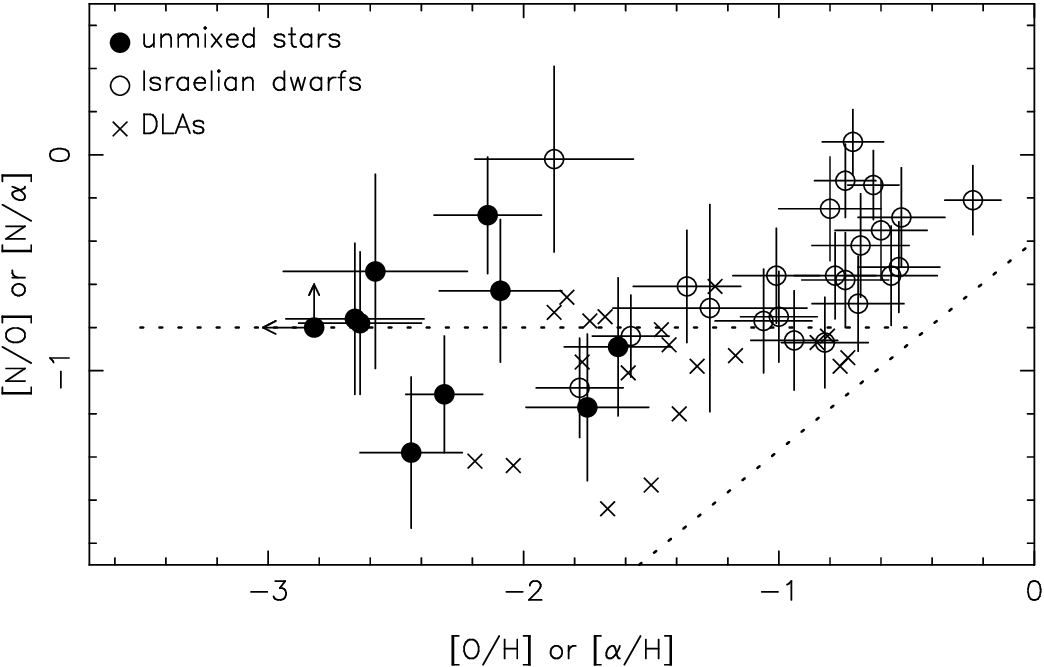} }
\caption {
Light-element abundance ratios in metal-poor dwarfs (Israelian et 
al.  \cite{IER04}, open circles), in our unmixed giants 
(filled circles), and in DLAs (crosses).  [N/O] vs.  [O/H] is
shown for the stars, ${\rm [N/\alpha]~ vs.~ [\alpha/H]}$ for the DLAs,
following Centuri\'on et al.  (\cite{CMV03}), where $\alpha$ may
denote O, S, or Si in different DLAs. The DLAs with ${\rm [\alpha/H]<-2}$
have ${\rm [N/\alpha] \approx -1.4}$, while the most metal-poor stars
have ${\rm [N/\alpha] \approx -0.8}$; but given the scatter in both
samples, this difference may not be significant.  The
horizontal and sloping dotted lines show the relations for primary and
secondary N production, respectively.  Neither sample shows any
dependence of N abundance on metallicity for [O/H] or ${\rm[\alpha/H] < 
-1.0}$ dex; thus, early N production was mainly primary.}
\label {NDLA}
\end {center}
\end {figure}

Fig. \ref{NDLA} shows [N/O] vs. [O/H] for our unmixed XMP giants, the 
metal-poor dwarfs studied by Israelian et al. (\cite{IER04}), and
the DLAs studied by Centuri\'on et al. (\cite{CMV03}) and Molaro et al. 
(\cite{molaro03}). Because O abundances have been measured in very few DLAs, we 
plot ${\rm [N/\alpha]}$ vs. ${\rm[\alpha/H]}$ for the DLAs (where $\alpha$ is O, 
S, or Si, depending on the DLA), which should be equivalent to [N/O] vs, [O/H] 
(Molaro \cite{molarocno}). Error bars have been omitted for the DLAs, but the ({\em internal}) error of [N/$\alpha$] in DLAs is claimed to be of order 0.02 
dex. 

In addition to the internal errors, systematic uncertainties affecting the two 
sets of observations include:\\
{\em (i)} The stellar data are subject to the neglect of 
3D effects in our 1D atmospheres; using 3D models (see section \ref{AbCandO}) 
would reduce the O abundance by typically $\sim$0.2~dex, but without a full 3D 
computation for the NH molecule the net effect on [N/O] is difficult to assess. 
In addition, our absolute N abundances also depend on the uncertain physical 
parameters for the NH band (see section \ref{azot1}) and were corrected by --0.4 
dex to force agreement with abundances from the CN band. Systematic offsets in 
our N/O ratios can therefore neither be excluded nor assessed quantitatively at 
present. New, accurate values of gf and dissociation energy for NH as well as 
studies of 3D model atmosphere effects on the band strength are urgently 
needed.\\ 
{\em (ii)} The DLA observations are affected by uncertain dust corrections. In 
particular, Si is easily locked onto grains, and the DLAs for which Si was 
used to represent O are subject to larger uncertainties.

For stars of higher metallicity ([O/H]$>-1.8$), the 
significant slope of [N/O] vs. [O/H] is indicative of secondary N
production (Israelian et al. \cite{IER04}). While noting the
possible systematic errors, we find most of our unmixed stars to have
[N/O]$\sim -0.9$, similar to the metal-poor stars of Israelian et al.
(2004) but extending to much lower [O/H] values.  This seems to agree
also with the main [N/$\alpha$] plateau of the DLAs.  However, while
the DLA data hint at the existence of two distinct plateaus, the stars
also occupy the region between them.  Note also that none of our stellar
[N/O] ratios is below the ``lower'' [N/$\alpha$] plateau of the DLAs.

The similarities and differences between our stellar sample and the DLAs can be 
summarised as follows:\\

\noindent Similarities:
\begin{enumerate}
\item Both stars and DLAs form a well-populated plateau at [N/O]$\sim -0.9$, 
independent of metallicity;
\item No star or DLA is found with [N/O]$ < -1.5$.
\end{enumerate}

\noindent Differences:
\begin{enumerate}
\item The stellar data extend to lower metallicity than DLAs, with [N/O]$\sim 
-0.9$ at the lowest values of [O/H].
\item Even when excluding the N-rich stars (as in Fig. \ref{NDLA}), [N/O] in 
stars scatters  more than in DLAs.
\end{enumerate}
Given the scatter of [N/O] in both stars and DLAs, these differences may not be 
significant, however.

Given the still-limited amount of data, it may be premature to attempt to fit 
all of these facts into a simple and coherent picture. We believe, however, that 
the lack of any significant trend of [N/O] with metallicity in both stars and 
DLAs requires primary N production in both. In Fig. \ref{NDLA} the dotted lines 
indicate the relations for primary and secondary N production from Centuri\'on 
et al. (\cite{CMV03}). For ${\rm [O/H] < -2}$, the mean value of [N/O] is close 
to the primary production line, which suggests that massive stars have enriched 
the ISM in N before our very old XMP stars and DLAs were formed. 

There is, however, no conclusive evidence whether the main site of this primary 
N production is massive Pop III stars (exploding as SNe II), AGB stars, or AGB 
supernovae (SN I.5). On the one hand, the positive correlation of [N/H] with 
[O/H] suggests an origin in massive stars, despite the large scatter in [N/O]. 
On the other hand, the large spread in [N/O] among Galactic stars might suggest 
that early AGB stars could play an important role. It may well be that N is 
produced in comparable amounts in {\em both} massive stars and AGB stars.

\section {Conclusions}
\subsection {Internal mixing}
Our careful analysis of CNO and Li abundances in extremely metal-poor giants 
from high-quality spectra has demonstrated that they fall cleanly in two groups:

\begin{itemize} 
\item The first group shows clear signs of mixing of significant amounts of 
CN-cycle products to the surface, very similar to the extra mixing found by 
Gratton et al. (\cite{GSC00}) in moderately metal-poor field giants. Evidence 
for this includes the non-detection of Li and low [C/N] in stars with low 
log $g$ values: these 
stars belong to the upper RGB or the HB. No signature of ON cycle 
processing is detected in the mixed stars. A forthcoming paper will discuss the 
${\rm ^{12}C/^{13}C}$ ratio to further study the extent of mixing in these 
stars.

\item The second group shows no signature of mixing with CNO burning layers: 
${\rm N_{Li} > 0.14}$ and ${\rm [C/N] > - 0.5}$. These stars have higher log $g$ and 
lie on the lower RGB. They may still have experienced the first dredge-up, 
but this is expected to change their atmospheric C and N abundances 
very little (Gratton et al.  \cite{GSC00}), or not at all (Denissenkov 
and Weiss \cite{DW04}).
\end{itemize}

The mixed and unmixed stars have approximately the same value of [(C+N)/Fe] 
$\approx +0.25$ dex, indicating that the difference between the two groups 
could be entirely due to internal mixing with CN-processed material. We 
thus confirm the results on extra-mixing by Gratton et al. (\cite{GSC00}) 
and extend them towards lower metallicity.

\subsection {The unmixed subsample}

We find that the C and N abundances in the {\em unmixed} XMP stars should
be close to those of the gas from which they were formed. These abundances thus 
provide useful constraints on the yields of the first stars (Pop III supernovae,
early AGBs, AGB supernovae,...)  and on any trends of these yields
with metallicity and perhaps time.

\begin{itemize}
\item The [C/Fe] ratio is remarkably constant with [Fe/H] and shows
only moderate scatter.  [C/Mg] and [C/O] decrease slightly with
metallicity from solar values at [Mg/H] $\approx -3.5$ and [O/H]
$\approx -3.0$, respectively.  We thus confirm the trends suggested by
Akerman (\cite{ACN04}) and extend them to even lower metallicity.\\

\item The N abundances show large scatter, and the sample of unmixed stars 
is limited in size; the abundance trends are therefore difficult to verify 
and interpret. While the mean values of both [N/Fe] and [N/Mg] appear to 
decrease with increasing metallicity (Fig. \ref {N-N-notmix}), this 
could be just a statistical effect.  In any case, the observed
trends are incompatible with secondary N production in the early Galaxy.

Our sample of unmixed XMP stars shows ranges of ${\rm -1.1<[N/Mg]<-0.0}$ and
${\rm -0.8<[N/Fe]<+0.2}$ (Fig. \ref {N-N-notmixb}).  Two tentative
interpretations of this large spread are suggested:\\
{\em i)} The primary N production by SNe II is close to the lower value of
[N/Mg] (--1.1), and stars with higher [N/Mg] values would be formed from 
material enriched by the winds of massive AGB stars; or \\
{\em ii)} The primary N production by SNe II is close to the higher value of
[N/Mg] (0.0), explaining the [N/Mg] ratio in most of our XMP stars.  
However, this primary production depends critically on various parameters and 
could be less active in some cases, especially in less massive supernovae. 
Some stars would then become "nitrogen-poor".

In order to test these different interpretations of the trends of
[N/Mg] and [N/Fe] vs.  metallicity, it would be important to measure
the N abundances in a larger sample of XMP stars (especially for ${\rm
[Fe/H]<-3.4}$ or ${\rm [Mg/H] < -3.2}$) to ascertain whether the small
spread of [N/Fe] and [N/Mg] at very low metallicity is real or just a
statistical effect.\\

%FIG 18
\begin {figure}
\begin {center}
\resizebox  {8.0cm}{4.75cm} 
{\includegraphics {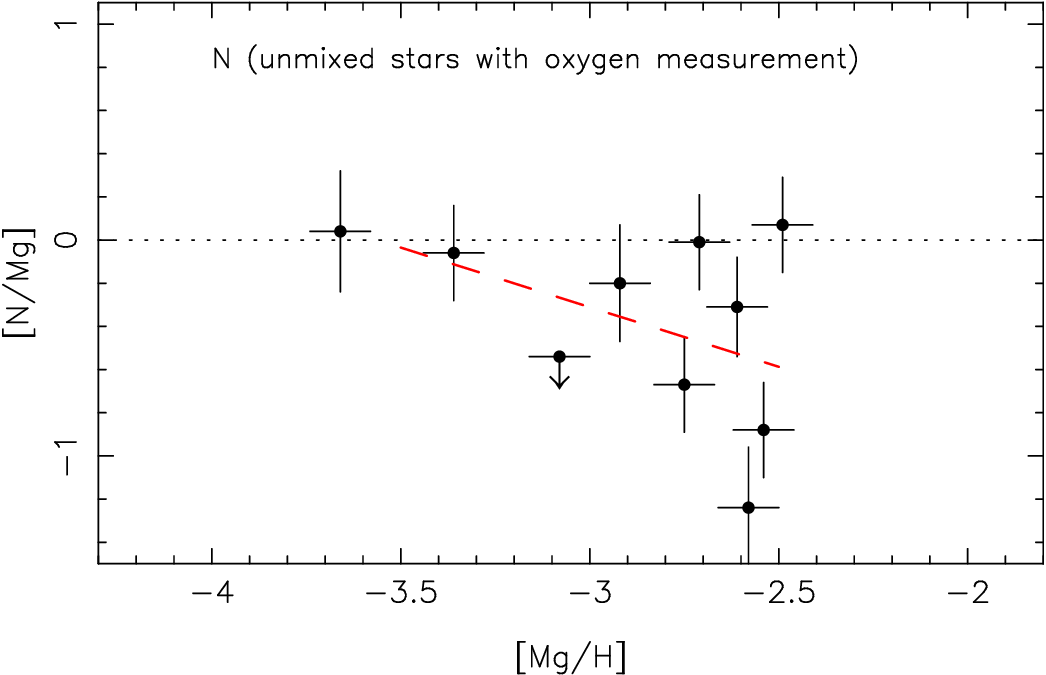} }
\caption {Same as Fig. \ref {N-N-notmix}, but only for stars included in 
Fig. \ref {NDLA}. The general decrease of [N/Mg] with [Mg/H] and the different 
behaviors of [N/Mg] and [N/O] remain visible.
}
\label {N-N-notmixwitho}
\end {center}
\end {figure}

\item A diagram of [N/O] vs. [O/H] for the same stars shows no 
visible trend in the range ${\rm -2.8 < [O/H] < -1.6}$, but the dispersion is 
very large ($\sigma=$ 0.31). The different behaviour of [N/Mg] and [N/O] is 
surprising, because both O and Mg are thought to be $\alpha$ elements. Because O 
was measurable in fewer stars than Mg, we have repeated the comparison using 
only stars with both Mg and O abundances (Fig. \ref{N-N-notmixwitho}). As seen, 
the slope of [N/Mg] versus [Mg/H] remains visible. Although the sample is small, 
the different trends in [N/Mg] and [N/O] seem hard to explain as just a 
statistical effect; they are are probably real and do not favor secondary 
production of N.\\

\item We have compared the ${\rm [N/O]}$ ratios in our XMP stars with ${\rm 
[N/\alpha]}$ data for DLA systems. The metallicity ranges of the two samples are 
complementary. In spite of the large scatter, the weight of the evidence 
suggests that the N production was primary, especially in the first phases of 
Galactic evolution. It cannot yet be decided, however, whether this primary N 
production was due to massive Population III supernovae, AGB stars, or perhaps 
AGB supernovae (SN I.5); all three sources may play significant roles.

\end{itemize}

The absolute values of the measured abundances are susceptible to
various uncertainties, from inaccuracy of the molecular parameters to
the effects of simplified models (LTE and 1D approximations).
However, these systematic effects should be similar from star to star
and the derived trends remain robust.  It is important, however, to
verify the effects of 3D NLTE model effects and improved molecular
data on the derived abundances as soon as possible.

As another item for the future, we have found that the abundance
anomalies observed in the ``mixed'' stars are due to mixing between
their surfaces and CNO-processing layers, similar to the scenario
(extra mixing) inferred by Gratton et al.  (\cite{GSC00}) and
theoretically explained by, e.g. Charbonnel (\cite{Cha95}) and
Denissenkov \& VandenBerg (\cite{DVB03} and references therein).
However, we cannot completely rule out the possibility that the
observed anomalies could result from pollution by AGB binary
companions.  It would thus be interesting to search for any companions
of the mixed stars (radial velocities, astrometric orbits, or
interferometry) in order to check this alternative.

\begin {acknowledgements}
We thank the ESO staff for assistance during all the runs of our Large
Programme.  T.C.B. acknowledges partial funding for this work from grants AST 
00-98508 and AST 00-98549, as well as from grant PHY 02-16783: Physics Frontiers 
Center/Joint Institute for Nuclear Astrophysics (JINA), awarded by the U.S. 
National Science Foundation. BN and JA thank the Carlsberg Foundation and the 
Swedish and Danish Natural Science Research Councils for partial financial 
support of this work.
\end {acknowledgements}

\end{document}